\documentclass[aos]{imsart}

\RequirePackage{amsthm,amsmath,amsfonts,amssymb}
\usepackage{graphicx}
\usepackage{enumerate}
\usepackage{multirow}
\usepackage{makecell}
\usepackage{url}
\RequirePackage[authoryear]{natbib}

\RequirePackage[OT1]{fontenc}
\RequirePackage{amsthm}
\RequirePackage[colorlinks]{hyperref}
\usepackage{graphicx}
\usepackage{ifthen}
\usepackage{amsfonts}
\RequirePackage{xcolor}
\usepackage{xspace}

\numberwithin{equation}{section}
\numberwithin{figure}{section}
\numberwithin{table}{section}
\theoremstyle{plain}
\newtheorem{thm}{Theorem}[section]

\theoremstyle{remark}

\newtheorem{remark}{Remark}[section]

\newcommand{\Prob}[1]{{\mbox{Pr}}\left(#1\right)}
\newcommand{\Unif}[1]{{\mbox{Unif}}\left(#1\right)}

\newcommand{\eop}{\hfill{\rule{0.05in}{0.1in}}}

\newcommand{\dsm}[1]{{\mbox{}}\left\{#1\right\}}
\newcommand{\iDSM}{{\mbox{{\it Interval}-DSM}}\xspace}
\newcommand{\sDSM}{{\mbox{{\it Simplex}-DSM}}\xspace}
\newcommand{\dDSM}{{\mbox{{\it Dirichlet}-DSM}}\xspace}

\startlocaldefs

\endlocaldefs

\begin{document}

\begin{frontmatter}
\title{A New Method for Multinomial Inference using Dempster-Shafer Theory}
\runtitle{Multinomial Inference using DS Inference}

\begin{aug}
\author[A]{\fnms{Earl C.}~\snm{Lawrence}\ead[label=e1]{earl@lanl.gov}},
\author[A]{\fnms{Alexander C.}~\snm{Murph}\ead[label=e2]{murph@lanl.gov}},
\author[A]{\fnms{Scott A.}~\snm{Vander Wiel}\ead[label=e3]{scottv@lanl.gov}}
\and
\author[B]{\fnms{Chaunhai}~\snm{Liu}\ead[label=e4]{chaunhai@purdue.edu}}
\address[A]{Statistical Sciences, Los Alamos National Laboratory\printead[presep={,\ }]{e1,e2,e3}}

\address[B]{Department of Statistics, Purdue University\printead[presep={,\ }]{e4}}
\end{aug}

\begin{abstract}
A new method for multinomial inference is proposed by representing the
cell probabilities as unordered segments on the unit interval and
following Dempster-Shafer (DS) theory.  The resulting DS
posterior is then strengthened to improve symmetry and learning
properties with the final posterior model being
characterized by a Dirichlet distribution.  In addition to
computational simplicity, the new model has desirable invariance properties related to category permutations, refinements, and coarsenings. Furthermore, posterior inference on relative probabilities amongst certain cells depends only on data for the cells in question. Finally, the model is quite flexible with regard to parameterization and
the range of testable assertions.  Comparisons are made to
existing methods and illustrated with two examples.
\end{abstract}



\end{frontmatter}

\section{Introduction}
Suppose that observed categorical data $\{X_1,\ldots,X_n\}$,
\[
         X_i \in \mathbb X=\{1,\ldots,K\} \qquad(i=1,\ldots,n)
\]
are modeled as independent over $i$ with 
\begin{equation}
\Prob{X_i=k|\theta} = \theta_k\qquad (k \in \mathbb X, \theta\in\Theta)
\label{eq:theSamplingModel}
\end{equation}
for unknown category probabilities $\theta=(\theta_1,\ldots,\theta_K)$, where
$\Theta$ is the $K-1$ simplex
\begin{equation}
\Theta = \{(\theta_1,\ldots,\theta_K):\;
\mbox{$\sum_{k=1}^K \theta_k=1,\, \theta_k\geq 0$ for all $k=1,\ldots,K$}\}.
\label{eq:K-1simplex}
\end{equation}
The problem is to infer $\theta$ from the observed data\footnote{Note that $n$ refers to the number of trials in the multinomial model, not to the number of data sets observed from such a multinomial model.}.

Inference on the multinomial distribution is a
challenging problem in statistics with a rich history. For the simplest case, {\it i.e.},
the Bernoulli model ($k=2, n=1$), the history traces back to \cite{bayes1958}, whose posterior, in modern statistics terms, is obtained by
assuming a uniform prior on the probability of {\it success}.  For the case of extending Bayes' methods to the more complicated binomial model ($k=2, n>1$), \cite{pearson1920} called this seemingly simple problem {\it the fundamental problem of practical statistics}.
For a review of several modern approaches to the binomial problem that compares each method's frequentist properties, see \cite{brown2001}.  

Research into the general multinomial model enjoys a similarly rich history (see, for instance, \cite{papoulis1965random}) and remains a heavily researched topic to this day.  In \cite{liu2009}, a likelihood ratio goodness-of-fit test is developed using a tubular neighborhood around the true multinomial model.  The authors in \cite{bu2020} develop a method for determining an ``optimal design" when modeling multinomial responses via maximizing the determinant of the Fisher information matrix.  These authors also provide a survey of several modern applications of modeling multinomial data.

There are many alternatives to standard Bayesian
methods for performing multinomial inference, such as the {\it imprecise Dirichlet model} (IDM) proposed by \cite{walley1996} (also see \cite{bernard2005}), and the generalized fiducial approach developed in \cite{hannig2016, liu2016}.  For the binomial model, the method of \cite{dempster1966} is
perhaps the best available because its inferential procedures (such as the credible interval) have good frequentist properties.  However, his extension to the multinomial model was seldom used -- partly because of its computational complexity -- until the development of the sampling method in \cite{jacob2021}.  The sampling method in \cite{jacob2021} uses linear programming to speed up sampling on a simplex.  While this method has made the model in \cite{dempster1966} (which requires such simplex sampling) more feasible, the experiment in \cite{hoffman2021} shows that the overall computational demands remain stark, and that \cite{dempster1966}'s model still does not scale well to higher dimensions.

In this paper, we develop an alternative Dempster-Shafer model (DSM) called the \dDSM that has many desirable properties, including those that \cite{walley1996} considered important for
multinomial inference, and significant computational advantages over the scheme introduced in \cite{jacob2021}.  Additional discussion on IDM is given in
Section~\ref{sec:real-data-example}; computational comparisons are made in Section \ref{sec:computational_advantages}.

In DS calculus, probability mass is allocated to sets or intervals rather than to individual events in a probability space.  This can be a major advantage over classical probability theory.  For instance, in applications where evidence supports not one but several events, a proportionate amount of probability can be assigned to the set of these events, rather than relying on mechanisms ignorant of data, such as the Principal of Insufficient Reason (see \cite{sentz2002}), to divide this evidence among several events.  Perhaps due to these advantages, DS Calculus continues to see use in modern applications \citep{edlefsen2009,zhao2022}.

In the \dDSM, the unit interval
$\mathbb U = [0, 1)$ is partitioned by the elements of $\theta$
arranged in an unknown order.  The resulting DS posterior for
$\theta$ is characterized by a Dirichlet distribution, making numerical computation simple in contrast to the \sDSM.  The \dDSM possesses the
three desirable principles of \cite{walley1996}, namely, {\it embedding},
{\it symmetry}, and {\it representation invariance}, as well as the
neutrality property of \cite{connor1969}.  It also
demonstrates the flexibility of a DS model with regard to point
assertions and alternative parameterizations.  In this sense, it brings
together the best aspects of IDM and of the DS framework for inference.

The rest of the paper is arranged as follows.  We begin with an overview of DS theory in Section \ref{sec:DStheory}.  Section
\ref{sec:interval} discusses a simple \iDSM in which $\theta$
partitions $[0,1)$ and uniform variables on $(0,1)$ convey
uncertainty.  Section \ref{sec:dirc} introduces an unknown ordering of
the partition elements to derive the new \dDSM for multinomial
inference in the case $n=1$.  Section \ref{sec:generaln} considers the
case of general $n$ and presents the main result (Theorem
\ref{thm:posterior01}) on the posterior model.  Section
\ref{sec:properties-ddsm} details some of the properties of the new
model.  Section \ref{sec:examples} compares inference using \dDSM and
its competitors through illustrative examples for a case with $\theta$
unconstrained in the $K-1$ simplex and a second case with $\theta$
constrained to a given lower dimensional manifold.  In Section \ref{sec:computational_advantages}, we show how \dDSM is computationally more efficient than its competitors. Section
\ref{sec:discussion} concludes the paper with a few remarks.

\begin{remark}
 It is worth noting that several of the papers cited above themselves reference an earlier version of this paper.  Although this earlier version was never published (for several reasons unrelated to the peer-review process), it continues to be cited by those interested in multinomial inference, even over a dozen years after the pre-print was released (see \cite{hannig2016, jacob2021, hoffman2021, lawrence2021, hoffman2022, fan2023, hoffman2024}).  The continued interest in this original work motivated this version of the paper, which is a meant to be a modern update to the unpublished draft that will bring this work to the attention of a broader readership.

While much of this paper is identical to the unpublished draft, we have added comparisons to a modern method that was developed in the last few years and have attempted to simplify and expand upon some of the arguments.  We also provide a further discussion on a theoretical paradox, referred to as the \textit{restricted permutation paradox}, that existed in the original formulation of the \dDSM.  We will show that the \dDSM continues to have great inferential potential at low computational cost, and that it deserves to be center-stage in future discussions on this topic. 
\end{remark}

\renewcommand{\oplus}{\;\dot{+}\;}
\renewcommand{\ominus}{\;\dot{-}\;}
\renewcommand{\S}{T}

\section{An overview of Dempster-Shafer Calculus} \label{sec:DStheory}
As a gentle introduction to DS theory, and as a means to see how probability may derived for an individual event from probability masses attributed to sets of events, we first review the fundamentals.  Define the {\it state space} as the space of all possible scenarios: the product of the space for all possible observed data and the
space for possible parameters, $\mathbb S = {\mathbb X}^n \times \Theta$.  For simplicity of notation in the subsequent, assume that $\mathbb S$ is countable.  Next, define a collection of {\it focal elements}, $\mathbb E \subset 2^{\mathbb{S}}$, each element of which is a {\it subset} of the state space, and a {\it basic probability assignment} (bpa) function ${\cal M}: \mathbb E \to [0,1]$.  Here, $2^{\mathbb{S}}$ denotes the power set of $\mathbb{S}$.  The bpa follows two rules that would also be followed by a probability measure on $2^{\mathbb{S}}$: 
\begin{equation} \label{eq:bpa_rules}
    \mathcal{M} (\emptyset) = 0; ~~~~~~~~~~~ \sum_{A \in 2^{\mathbb{S}}} \mathcal{M}(A) = 1.
\end{equation}
The bpa, also called the belief function, is a generalization of a probability measure.  Note that \eqref{eq:bpa_rules} \textit{does not} require that probability assigned to a focal element imply that any probability is assigned to the strict subsets of that focal element.  This is an explicit breech of the Additivity Property required in probability theory and exhibits a central theme of DS theory: that evidence for a set of possible events need not also be explicitly divided among each individual event in that set.  Instead, to make probability statements on an individual event, potentially-conflicting or overlapping evidence is combined in a way that reflects uncertainty.  For a full axiomatic development of the bpa, see \cite{shafer1978, dempster2008}.  

Once a complete DSM is established, denoted using the three critical components $\dsm{\mathbb S,\, \mathbb E,\, {\cal M}}$, evidence can be combined to make statements on any assertion of interest $A\subset \mathbb{S}$, where $A$ can be an event or a set of events.  In DS theory, evidence for and against the assertion $A$ are derived from the basic probability allocations from $\mathcal{M}$, and there are several common derivations used to understand the validity of $A$.  The probability $p$ for the truth of $A$, sometimes referred to as \textit{Belief}, is the sum of allocations for every subset of $A$:
\begin{equation} \label{eq:belief}
    \text{Bel}(A) = p = \sum_{B: B\subseteq A} \mathcal{M}(B).
\end{equation}
Belief is the sum of all evidence that support \textit{only} some or all of the events in $A$.  Meanwhile, the probability $q$ against the truth of $A$ is this same quantity for the complement of $A$ in $2^{\mathbb{S}},$ denoted by $\overline{A}$.  That is,
\begin{equation} \label{eq:belief_against}
    q = \sum_{B: B\subseteq \overline{A}} \mathcal{M}(B),
\end{equation}
which should be interpreted as the sum of all evidence that support \textit{only} events that are not in $A$.  Note that allocations given to sets that contain some events in $A$ and some events not in $A$ (events $B$ such that $A\cap B \neq \emptyset$) are not included in the calculation of \eqref{eq:belief} \& \eqref{eq:belief_against}.  The sum $r$ of these allocations,
\begin{equation}
    r = \sum_{B: B \not\subset A, B \cap A \neq \emptyset } \mathcal{M}(B),
\end{equation}
is seen as evidence that might be for or against $A$, and is colloquially termed ``don't know" \citep{dempster2008}.  Since this evidence could ``go both ways," it can be combined with $p$ to quantify the amount of all possible evidence for the truth of $A$, called the \textit{plausibility} for the truth of $A$ (see, {\it e.g.}, \cite{shafer1976}).  The plausibility of $A$ can equivalently be written as,
\begin{equation*} 
    \text{Pl}(A) = p + r = \sum_{B: B \cap A \neq \emptyset} \mathcal{M}(B).
\end{equation*}
For this paper, we focus on the three derivations $(p,q,r)$ as the primary DS output and informative quantities for inference on the multinomial model.

Let $\mathbf{e}$ be a random focal element in $\mathbb{E}$ that is distributed according to the probability allocations determined by $\mathcal{M}$.  In this paper, we will find it occasionally useful to view $\mathbf{e}$ as a function of some auxiliary random variable that does not depend on the data nor the parameters: $\mathbf{e}(U)$ (see \cite{liu2008}).  For notional convenience, a random element from a DSM will be denoted $\mathbf{e}\sim \mathcal{M}$.

Information from independent DSMs $\mathbf e_1 \sim {\cal
M}_1$ and $\mathbf e_2 \sim {\cal M}_2$ defined on the same state space can be combined according to several different combination rules.  The original combination rule, called the \textit{Dempster's rule of combination} \citep{dempster1972}, combines evidence from both sources whenever they agree completely and ignores all evidence that conflicts in any way, and is similar to combining probabilities via a product measure \citep{martin2010}.  For a further explanation of how to combine DSMs, and an overview of different combination rules, see \cite{sentz2002}.  In this paper, we will focus primarily on Dempster's rule of combination.  Inference on $\theta$ from a DSM $\mathbf e \sim \cal M$ with $\mathbf e \subseteq {\mathbb X}^n
\times \Theta$ is obtained by conditioning on an observed set of data, projecting the
random set $\mathbf e$ onto the space $\Theta$, and then computing
$(p,q,r)$ for assertions $A\subset\Theta$.

For the multinomial model, \cite{dempster1966} proposed a DSM based on uniform sampling of $\theta$ from the $K-1$ simplex (he called this a sampling structure of the {\it second kind}). We call this model the \sDSM.  It has some unappealing features detailed in Section \ref{sec:properties-ddsm} and that seem to arise from the difficulty in predicting a uniformly distributed point in the $K-1$ simplex.  Furthermore, even with recent computational developments (see \cite{jacob2021}), it suffers from high computation complexity for larger values of $K$ (see Section \ref{sec:computational_advantages}).

\cite{walley1996} IDM handles uncertainty by specifying a family of
Bayesian models for analysis of multinomial data.
Indeed, an IDM produces lower and upper probabilities for certain
assertions, such as $\{\theta_1 \leq \theta_0\}$ with a
fixed $\theta_0\in [0, 1]$. However, for point assertions such as
$\{\theta_1 = \theta_0\}$, these upper and lower probabilities are
both zero and thus IDM obtains $(p, q, r)=(0, 1, 0)$.  Contrast this with a DS
result for which $r > 0$, admitting the plausibility of the point
assertion. In this respect, random set probability models like DS are more
appealing than imprecise Bayesian models like IDM.
Furthermore, the combination operation of DS calculus provides a
simple way of restricting a saturated multinomial DSM to infer
a constrained multinomial parameter (see the example in Section~\ref{sec:real-data-example}).

\section{A Simple Interval DSM  for the Multinomial Model}
\label{sec:interval}

\begin{figure}[t]
  \centering
\includegraphics*[width=3.5in]{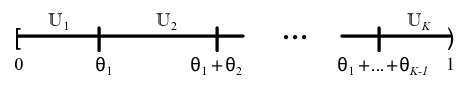}
  \caption{Partition of ${\mathbb U}=[0,1)$ into ${\mathbb U}_1, \ldots, {\mathbb U}_K$.}
  \label{fig:partition}
\end{figure}

The multinomial model
\eqref{eq:theSamplingModel}--\eqref{eq:K-1simplex} can be described as
follows.  Let the parameter $\theta=(\theta_1,\ldots,\theta_K)\in\Theta$
partition the unit interval ${\mathbb U}=[0,1)$ into disjoint segments
${\mathbb U}_1 = [0, \theta_1)$ and
${\mathbb U}_k = [\theta_1+\cdots+\theta_{k-1},\ \theta_1+\cdots+\theta_k)$
for $k=2,\ldots,K$.  Figure~\ref{fig:partition} illustrates the partition.

Each observation $X_i$ is associated
with a value $U_i\in\mathbb U$ through the multi-valued mapping
\begin{equation}
        X_i = a(U_i, \Theta) \qquad \text{where} \qquad a(U_i, \Theta) = \sum_{k \in \mathbb{X} } k * I(U_i \in \mathbb{U}_k),
\label{eq:first-kind}
\end{equation}
where $i \in \{1,\dots, n\}$ and $I(\cdot)$ is an indicator function.  As explained further in the next section, samples from the Multinomial distribution are obtained by assuming $U_i\stackrel{iid}{\sim} \Unif{0,1}$ {\it a priori} and using \eqref{eq:first-kind}; these sampling choices have an impact on inference for the parameters.

\subsection{A basic DSM built on the sampling model (\ref{eq:first-kind})} \label{sec:basic_DSM}
For notational convenience, take $n=1$ and write $X_1=X$ and  $U_1=U$ in \eqref{eq:first-kind} and note that the observation $X$ is determined by the scalar latent variable $U$.
Following \cite{liu2008}, we call $a$ in (\ref{eq:first-kind}) the {\it auxiliary equation} and $U$ the {\it auxiliary variable}.

Let $U^*$ denote the realized (but unobserved) value of $U$ associated
with observed data $X$.
DS inference on $\theta$ proceeds by ``inverting'' the auxiliary
equation $a$~\eqref{eq:first-kind} and ``continuing to believe''
(Dempster, 2008) that $U^*\sim\Unif{0, 1}$.
The phrase ``continuing to believe'' indicates a progression from an
{\it a priori} sampling model for the generation of $U^*$ to a
posterior belief about $U^*$ after it has been realized but
independent of any additional information ({\em e.g.}, the value of $X$).  As shown below, given that $U$ is distributed according to some \textit{pivotal} measure, $U \sim \mu$, inversion of $a$ defines a distribution of random sets in the state space.  Use of an auxiliary variable through an auxiliary equation to determine a distribution on the state space is more generally known as Fisher's fiducial argument \citep{fisher1935}.  DS theory is sometimes regarded as a generalization of Fisher's original argument that allows for random sets rather than requiring random points \citep{martin2010}.

Formally, the DSM (on 
$\mathbb{S} = \mathbb{X} \times \Theta \times \mathbb{U}$)
that represents the above ``continuing to believe'' model for $U^*$ 
has focal elements 
\begin{align}
    {\bf e}_1(U) &= \{(X,\theta,U):
    \; X\in \mathbb{X}, 
    \  \theta \in \Theta, 
    \  U^* = U
    \}
    \qquad
    (U \sim \Unif{\mathbb{U}}).
    \label{eq:IXthetaU}
\end{align}
The third coordinate represents $U^*$ and its stated distribution represents ``continuing to believe'' that $U^*\sim \text{Unif}(\mathbb{U})$, even after knowing that the value of $U^*$, though unobserved, has, in fact, been realized.
This DSM is {\em vacuous} with respect to $(X,\theta)$, meaning that its focal elements project to all of $\mathbb{X} \times \Theta$ with probability one:
\begin{align*}
    \mu \{ U : p_{1,2}(\mathbf e_1(U)) = \mathbb{X} \times \Theta \} = 1
\end{align*}
where $p_{i,j,\ldots}(\cdot)$ denotes projection of a set onto the specified coordinates.  Less pedantically, we say that Model~\eqref{eq:IXthetaU} assigns unit mass to the full space $\mathbb{X} \times \Theta$. As a consequence, it assigns zero mass to the collection of all proper subsets of $\mathbb{X} \times \Theta$.

The auxiliary equation for $a$ (\ref{eq:first-kind}) is represented by the DSM with a single unit-mass focal element
\begin{align}
   \mathbf e_2 &= \{(X, \theta,U^*):\; 
   X\in \mathbb{X},
   \ \theta \in \Theta, 
   \ U^* \in \mathbb{U},
   \ X = a(\theta, U^*)
   \} .
\label{eq:Ia-eqn}
\end{align}
This set contains all $(X,\theta,U^*)$ triples consistent with the auxiliary equation, recalling that the partition $\{{\mathbb U}_1,\ldots,{\mathbb U}_K\}$ in $a$ is a function of $\theta$. The set $\mathbf e_2$ is deterministic, not dependent on $U$, and therefore has unit mass with respect to $\mu$.

Independent combination of (\ref{eq:IXthetaU}) and (\ref{eq:Ia-eqn})
according to Dempster's rule of combination intersects $\mathbf e_1$ and $\mathbf e_2$.  Subsequent projection to $\mathbb{X}\times\Theta$ produces
\begin{align}
  \mathbf e(U) 
  &= p_{1,2}(\mathbf e_1 \cap \mathbf e_2) \nonumber \\
  &= \{(X, \theta): 
    \; X\in \mathbb{X}, 
    \ \theta\in\Theta, 
    \ X = a(\theta, U)
   \} 
   \qquad  (U\sim \Unif{\mathbb U}),
\label{eq:IXtheta}
\end{align}
the set of $(X,\theta)$ pairs that are consistent with the randomly drawn $U$.
Model~\eqref{eq:IXtheta} is the $\iDSM$ for multinomial
inference with $n=1$.
We reiterate that $\iDSM$ is specified from the auxiliary equation $a$ by ``continuing to believe'' that uncertainty in $U^*$ is appropriately modeled by its prior $\Unif{0, 1}$ distribution, even after learning that the value of $U^*$ has been realized.

Consider  model~\eqref{eq:IXtheta} conditional on a known
$\theta$ and, alternatively, on a known $X$.

\paragraph{1) Known $\theta$.}
Knowledge of $\theta$ is represented by the DSM with
a single unit-mass focal element projected to $\mathbb{X} \times \Theta$
\[
   \mathbf e_{3,\theta} 
   = p_{1,2}(\{(X, \theta, U):
   \; X\in\mathbb{X}, 
   \; U \in \mathbb{U} \}).
\]
Combining this with (\ref{eq:IXtheta}) and projecting to $\mathbb{X}$ gives
\begin{align*}
   \mathbf{e}_\theta(U)
   & = p_1(\mathbf e_{3,\theta}  \cap \mathbf e(U) ) \\
   & =
   \{X:\; X\in \mathbb{X},
   \; X = a(\theta, U)
   \}\qquad (U\sim \Unif{{\mathbb U}})
\    
\end{align*}
which is equivalent to (\ref{eq:first-kind}), the multinomial sampling model.

A DSM that preserves the postulated sampling model is called a
\textit{basic} DSM.  The key elements for constructing a basic DSM are
an auxiliary equation (like $a$ in \ref{eq:first-kind}), its corresponding DSM
(like \ref{eq:Ia-eqn}), and the DSM representing ``continuing to
believe" the original auxiliary variable's sampling distribution (like \ref{eq:IXthetaU}).

\paragraph{2) Observed $X$.}
Next consider conditioning on an observed value of $X$ to obtain the model for posterior inference.  An observation $X$ is represented by the DSM with a single unit-mass
focal element projected to $\mathbb{X} \times \Theta$
\[
\mathbf e_{4,X} = p_{1,2}(\{(X, \theta, U):
    \; \theta\in\Theta, 
    \; U \in \mathbb{U} \}) .
\]
Combining this with (\ref{eq:IXtheta}) and projecting to $\Theta$
gives
\begin{align}
 \mathbf e_X(U) 
    & = p_2(\mathbf e_{4,X}  \cap \mathbf e(U) ) \nonumber \\
    &= \{\theta:\; \theta \in \Theta, X = a(\theta, U)\} \nonumber \\
    &= \{ \theta:\; \theta \in \Theta, 
          \theta_1 + \cdots + \theta_{X-1} \leq U < \theta_1 + \cdots + \theta_X
        \}
\label{eq:Iposterior}
\end{align}
where $U\sim \Unif{{\mathbb U}}$.

Although the posterior on $\theta$ \eqref{eq:Iposterior} for \iDSM has
the advantage of being simple, it has the disadvantage of
depending on the order in which the $\theta_k$ partition $\mathbb U$.
This lack of symmetry is unappealing and led \cite{dempster1966} to
propose the \sDSM that preserves
the desirable symmetry about the categories. However, ``continuing to
believe'' in a $(K-1)$-dimensional auxiliary
vector for a single categorical observation $X$ leads to some unappealing properties for the \sDSM (see the Supplementary Materials \citep{lawrence2024}).

To avoid the effect of the order of $\theta_1,\ldots,\theta_K$ yet
retain the simplicity of a univariate auxiliary variable,
the \dDSM in the following section incorporates an unknown permutation into the auxiliary equation.

\section{The Dirichlet-DSM for \texorpdfstring{$n=1$}{n1}}
\label{sec:dirc}
Continue to partition the unit interval ${\mathbb U}=[0,1)$ into disjoint segments
$\mathbb{U}_1,$ \ldots, $\mathbb{U}_K$ with lengths $\theta_1$, \ldots
,$\theta_K$, such that $\sum_{i=1}^{K} \theta_i = 1$.  This time,
however, the ordering of the segments is assumed unknown.
An important result is that inference about
$\theta$ is possible without information on the order of the
segments.

\subsection{An unknown permutation}
\label{sec:an-unkn-perm}

\begin{figure}[t]
  \centering
\includegraphics*[width=3.5in]{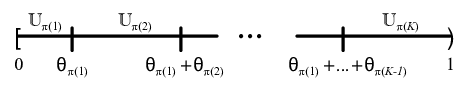}
  \caption{Partition of ${\mathbb U}=[0,1)$ into ${\mathbb U}_1,
    \ldots, {\mathbb U}_K$ ordered by a permutation $\pi$.}
  \label{fig:partitionPi}
\end{figure}

Let $\pi$ denote an unknown permutation of $\{1,\ldots,K\}$.
As before, $\mathbb{U}_k$ represents the interval of length $\theta_k$
but now it is redefined as
\begin{equation}\label{eq:Upi}
  \mathbb{U}_{k, \pi} =
  \begin{cases}
    [0,\ \theta_{k} ) &\text{if $\pi(1)=k$}, \\
    \left[\displaystyle\sum_{i=1}^{\pi^{-1}(k)-1} \theta_{\pi(i)},\
      \sum_{i=1}^{ \pi^{-1}(k) -1} \theta_{\pi(i)}+\theta_k\right)
    &\text{if $\pi(1) \neq k$}
  \end{cases}.
\end{equation}
Previously, each partition of $[0,1)$ was
a function of $\theta$; now it is a function of both $\theta$ and
$\pi$.  Figure~\ref{fig:partitionPi} illustrates the partition ordered
according to $\pi$.  Under this formulation the auxiliary equation is the same as in \eqref{eq:first-kind} but with ${\mathbb U}_k$ replaced by $\mathbb{U}_{k, \pi}$.  To make this clear in the subsequent, we denote the auxiliary equation $a_\pi$.
Again, $U \sim \Unif{0,1}$ results in multinomial sampling.

Developing a DSM for the updated auxiliary equation from \eqref{eq:first-kind} using \eqref{eq:Upi} follows the same steps used to develop \iDSM in Section \ref{sec:basic_DSM}.  As before, let $U^*$ denote the
realized (but unobserved) value of $U$ associated with observed data
$X$.  The DSM defined by ``continuing to believe" that
$U^*\sim\Unif{\mathbb U}$ is
\begin{align} \label{eq:PXthetapiU}
   \mathbf e_1(U) =& \{(X, \theta,\pi,U^*):\; X\in \mathbb{X}, \theta\in\Theta, \pi \in \Pi, U^*=U \} \qquad
   (U \sim \Unif{{\mathbb U}}), 
\end{align}
where $\Pi$ is the space of permutations.
A second DSM represents the auxiliary equation in \eqref{eq:first-kind} with a single element having unit mass:
\begin{align} \label{eq:Pa-eqn}
\mathbf e_2 = & \{(X, \theta,\pi,U^*):\; X\in \mathbb{X}, \theta\in\Theta, \pi \in \Pi, U^* \in \mathbb{U}, X = a_\pi (\theta, U^*) \}.
\end{align}
Independent combination of (\ref{eq:PXthetapiU}) and (\ref{eq:Pa-eqn})
involves intersections $\mathbf e_1$ and $\mathbf e_2$.  Subsequent
projection to $\mathbb{X}\times\Theta\times\Pi$
produces
\begin{align} \label{eq:PXthetapi}
  \mathbf e(U) 
  & = p_{1,2,3}( \mathbf e_1(U) \cap \mathbf e_2) \nonumber \\
  & =\{(X, \theta, \pi):\; X\in \mathbb{X}, \theta\in\Theta,  \pi \in \Pi, X = a_\pi (\theta, U) \} 
  \qquad  (U \sim  \Unif{{\mathbb U}}), 
\end{align}
recalling that $a_\pi$ is a function of both $\theta$ and $\pi$.

First consider model (\ref{eq:PXthetapi}) conditioned on known parameters $(\theta, \pi)$; then consider conditioning on observed data $X$.

\paragraph{1) Known $\theta$ and $\pi$.}
Knowledge of the parameters is represented by the DSM with a single unit mass focal element projected to 
$\mathbb{X} \times \Theta \times \Pi$
$$
\mathbf e_{3,\theta,\pi} = p_{1,2,3}(\{(X, \theta, \pi, U):
    \; X\in\mathbb{X}, 
    \; U \in \mathbb{U} \}).
$$
Combination with (\ref{eq:PXthetapi}) and projection to $\mathbb{X}$ gives
\begin{align*}
    \mathbf{e}_{\theta, \pi}(U)
    & =  p_1( \mathbf e(U) \cap \mathbf e_{3,\theta,\pi})  \nonumber \\
    & =\{X:\; X\in \mathbb{X}, X = a_\pi (\theta, U)   \}
    \qquad (U\sim \Unif{{\mathbb U}}),
\end{align*}
which is equivalent to the multinomial sampling model through the dependence of $a_\pi$ on $\theta$ and $\pi$.

\paragraph{2)  Observed $X$.}
An observation $X$ is represented by the DSM with a single unit mass focal element projected to 
$\mathbb{X} \times \Theta \times \Pi$
\begin{align} \label{eq:PthetapiX}
    \mathbf e_{4,X} 
    & = p_{1,2,3}\{(X, \theta, \pi, U):
    \; \theta\in\Theta,  
    \; \pi \in \Pi, 
    \; U \in \mathbb{U} \}).
\end{align}
Combination with (\ref{eq:PXthetapi}) and projection to $\Theta
\times \Pi$ produces the posterior DSM for $(\theta,\pi)$
\begin{align} \label{PthetapigivenX}
    \mathbf{e}_{X}(U) 
    & =  p_{2,3}( \mathbf e(U) \cap \mathbf e_{4,X}) \nonumber \\
    & = \{(\theta, \pi):\; \theta\in\Theta,  \pi \in \Pi, X = a_\pi(\theta, U) \} 
    \qquad (U\sim \Unif{{\mathbb U}}).
\end{align}

Figure~\ref{fig:sets} illustrates a single set $\mathbf{e}_{X}(U)$
for the trinomial model with $X=1$ and $U=0.75$.
All six permutations $\pi$ are enumerated across the top of the
figure and the corresponding regions in $\Theta$ are shown by red
polygons on the simplices based on the definition of ${\mathbb U}_X$
given in equation~(\ref{eq:Upi}).

\subsection{A restricted permutation}
\label{sec:restr-perm}

\begin{figure}[!t]
\centering
\rotatebox{0}{\includegraphics*[width=1.21in]{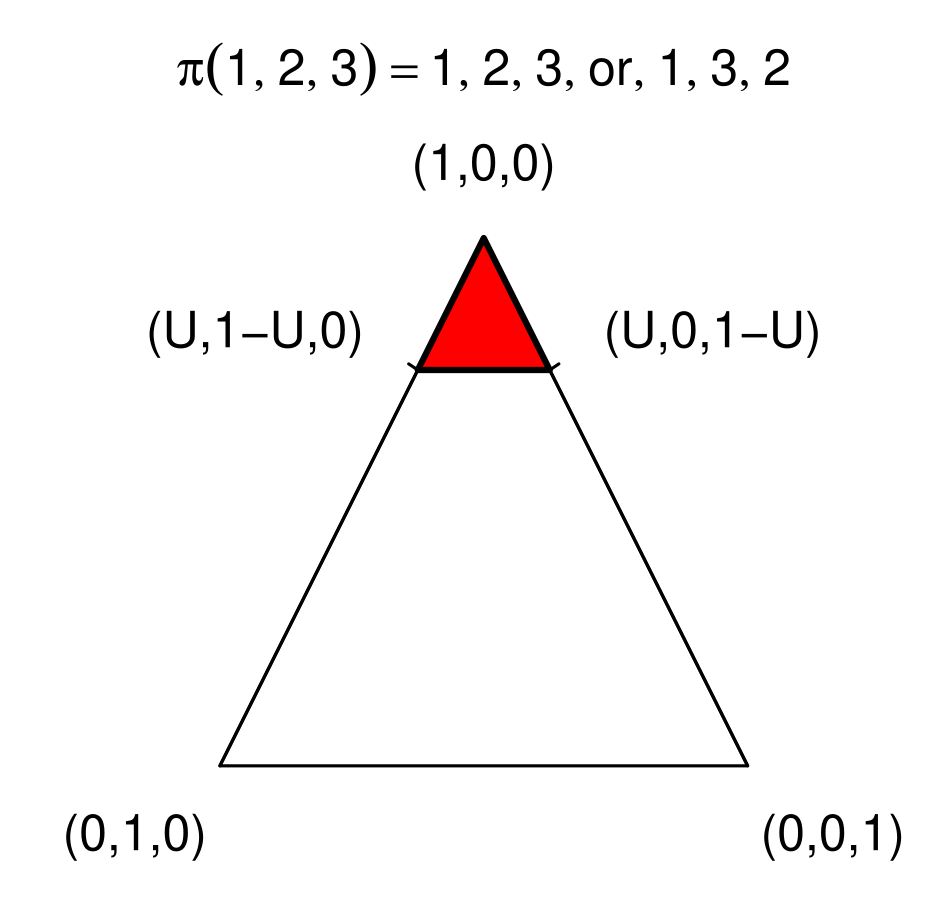}}
\rotatebox{0}{\includegraphics*[width=1.35in]{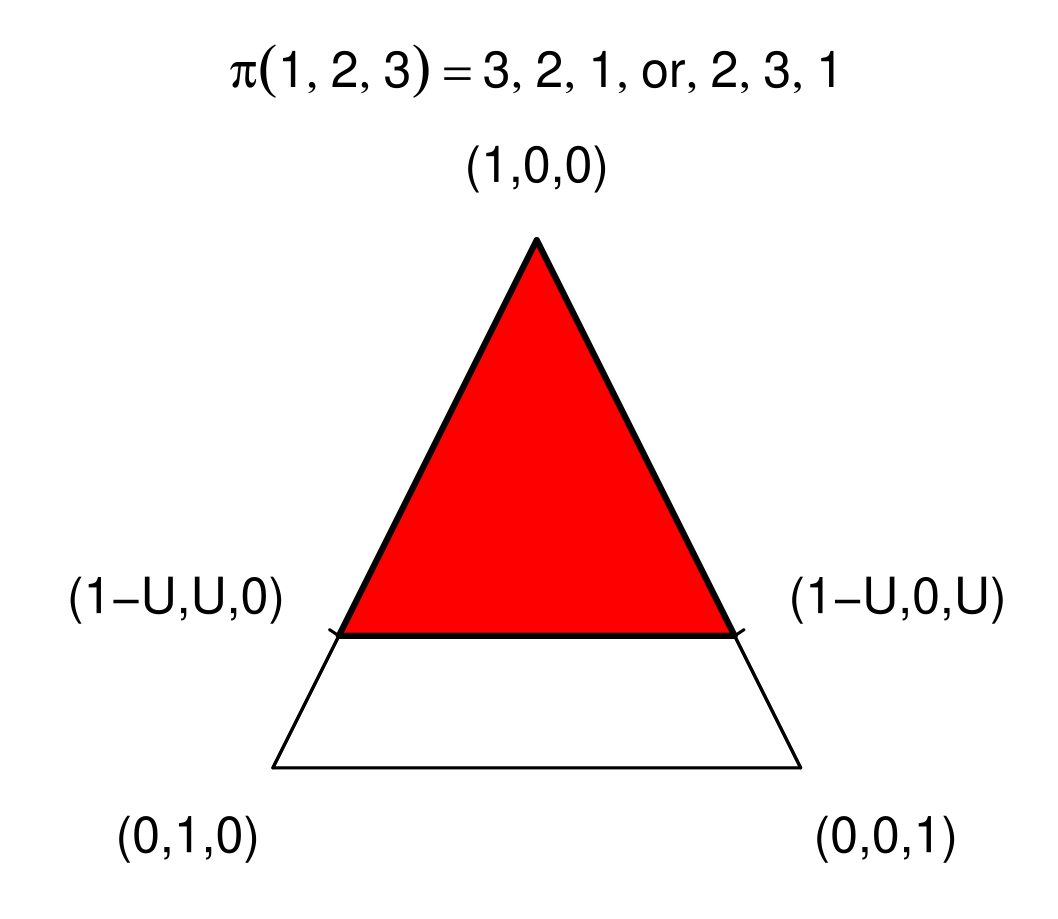}}
\rotatebox{0}{\includegraphics*[width=1.21in]{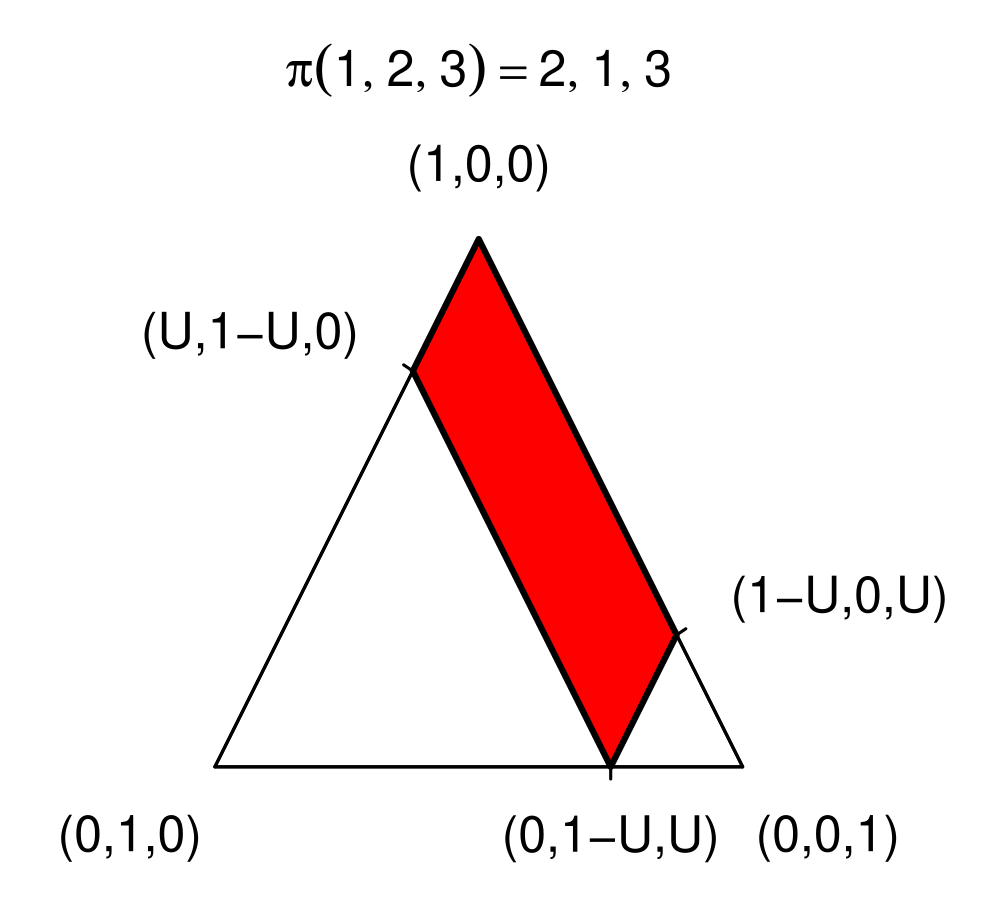}}
\rotatebox{0}{\includegraphics*[width=1.21in]{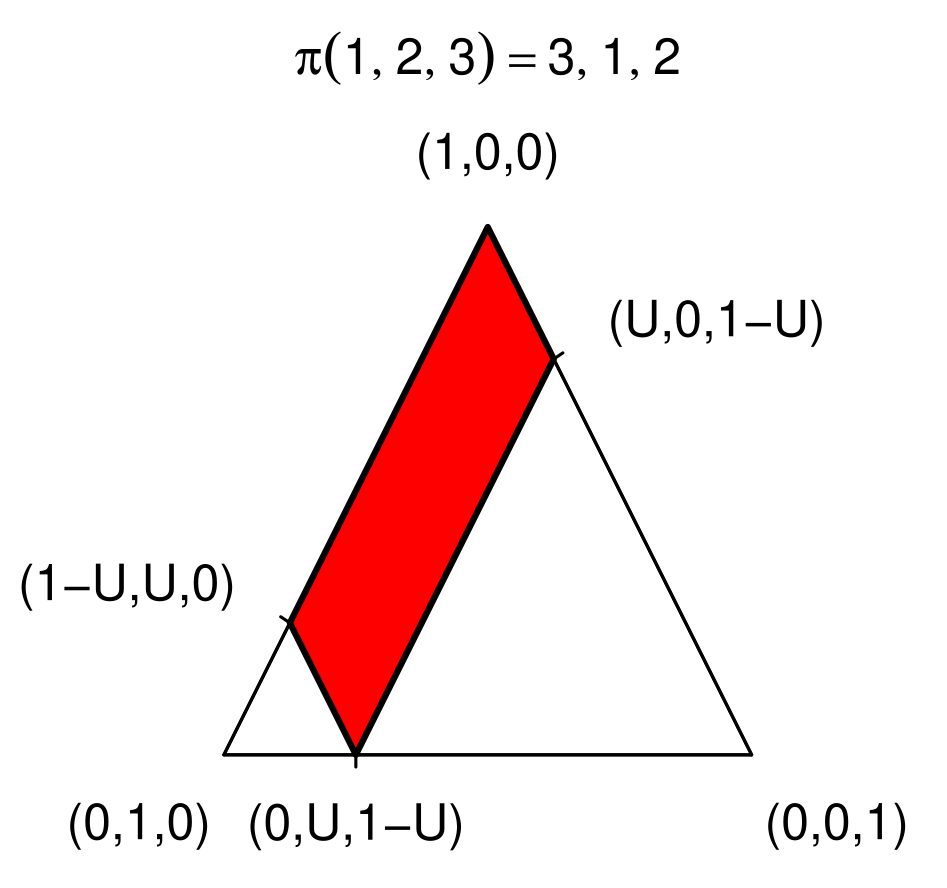}}
\caption{ A realization of the random set of the posterior
distribution for ($\theta, \pi)$ in the trinomial model.
The parameter space for $\Theta$ is shown by
the outer large triangle, representing the (standard) $3$-dimensional simplex.
The inner shaded regions represent the values of $\theta$ that satisfy the auxiliary equation for $U=0.75$ and $X$=1.
}
\label{fig:sets}
\end{figure}

As seen from Figure~\ref{fig:sets} projecting
$\mathbf{e}_{X}(U)$ onto $\Theta$ would produce a set that is {\em not}
vacuous with respect to the relative probabilities of the unobserved classes
\begin{equation} \label{eq:relativeTheta}
    \theta_{-X} \equiv 
    \left( \frac{\theta_1}{1-\theta_X},
  \ldots, \frac{\theta_{X-1}}{1-\theta_X},
  \frac{\theta_{X+1}}{1-\theta_X}, \ldots,
  \frac{\theta_{K}}{1-\theta_X} \right). 
\end{equation}
In particular, the
elements of $\mathbf{e}_{X}(U)$ illustrated in the two rightmost plots
limit the possible values of $\theta_{-X}$ within the $K-2$
dimensional simplex whenever $\theta_X$ is less than $U$.  This is
undesirable because it implies that the outcome $X$ provides
information about the relative probabilities $\theta_{-X}$.  As an example of this, consider the trinomial example with permutation $\pi(1,2,3) = 3,1,2$ (the scenario of the right-most plot in Figure \ref{fig:sets}) and suppose $\theta_1 = 0.6$.  Then $\theta_{-1} \equiv \left(2.5 \theta_2, 2.5 \theta_3 \right) = \left(\tilde \theta_2, \tilde \theta_3 \right)$.  Since $\theta_2 \in [0, 0.25)$, this implies that $\tilde \theta_2 \in [0, 0.625)$.  Thus $\tilde \theta_2$ is constrained by an upper bound that will change depending on the value of $\theta_1$.  Meanwhile, suppose $\pi(1,2,3) = 3,2,1$ (the scenario of the second plot in Figure \ref{fig:sets}) and $\theta_1 = 0.8$.  Then $\theta_{-1} \equiv \left(5 \theta_2, 5 \theta_3 \right) = \left(\tilde \theta_2, \tilde \theta_3 \right)$.  Since $\theta_2 \in [0, 0.2)$, $\tilde \theta_2 \in [0, 1)$ and is not constrained by $\theta_1$.  Note that $\tilde \theta_2$ and $\tilde \theta_3$ still depend on each other since $\tilde \theta_2 + \tilde \theta_3 = 1$ necessarily in both examples.

The implication of certain permutations -- that observation of category $X$  affects the relative probabilities among the remaining categories -- is too strong because these permutations impose structure on the unobserved probabilities beyond requiring that they exist on a simplex.  Therefore, we choose to modify
DSM~(\ref{eq:PXthetapi}) so as to exclude this extra by restricting to permutations that place $X$ either first or last---that is, either $\pi(1)=X
\text{ or } \pi(K)=X$, corresponding to the two
leftmost plots illustrated in Figure~\ref{fig:sets}.

Even the above restriction on $\pi$ is insufficient.  Consider
projecting the two leftmost plots of Figure~\ref{fig:sets} onto
$\Theta$ with the result being the union of the red regions.  Since $\theta_X$ can be as small as both both $U$ and $1-U$, this projection onto $\Theta$ puts a lower bound on $\theta_X$ of at most $1/2$
with an implication that even a large number of consecutive draws of
$X=1$ will provide no evidence {\em for} the assertion that $\theta_1
> 1/2.$ Thus, allowing both $\pi(1)=X \text{ and } \pi(K)=X$
represents a belief that is too weak with respect to $\theta_X$.  For
this reason we impose the further restriction that $\pi(1)=X$, represented as the unit mass DSM
\begin{equation} \label{eq:orderRestrictedDSM}
    \mathbf e^* 
    = \{(X, \theta,\pi,U):
    \; X\in \mathbb{X}, 
    \ \theta\in\Theta, 
    \ \pi \in \Pi, 
    \ U \in \mathbb{U}, 
    \ \pi(1) = X \}
    .
\end{equation}
Folding this into the definition of $\mathbf e(U)$~(\ref{eq:PXthetapi})
and then projecting onto $\mathbb{X}\times\Theta$ produces a simple
DSM for $(X,\theta)$:
\begin{align}
    \mathbf e^*(U) 
    & =  p_{1,2}( \mathbf e_1(U) \cap \mathbf e_2 \cap \mathbf e^*) \nonumber \\
    & = \{(X,\theta):\; X\in\mathbb{X}, \theta\in\Theta, \theta_X > U \}
    \qquad (U \sim  \Unif{\mathbb U}).
    \label{eq:ddsm}
\end{align}
We call model~(\ref{eq:ddsm}) the \dDSM (for $n=1$) for reasons that will become
clear in Section~\ref{sec:generaln}.

An attractive feature is that (\ref{eq:ddsm}) is vacuous
with respect to the relative probabilities, $\theta_{-X}$, in
the unobserved classes.  The choices we make involve weakening our inference on these unobserved cells, where a vacuous inference is as weak as possible.  Correspondingly, we have strengthened the inference on the observed category. 

Although the starting model~(\ref{eq:PXthetapi}) is a
basic DSM, the same is not true of the modified model~(\ref{eq:ddsm}).  In fact,
conditioning (\ref{eq:ddsm}) on a given $\theta$ and projecting to $\mathbb{X}$ gives
sets with a random number of elements:
\begin{align*}
    \mathbf{e}_{\theta}(U) 
    & = p_1( \mathbf e^*(U) \cap \{ (X,\theta): X \in \mathbb{X} \} ) \\
    & = \{X:
    \; X\in \mathbb{X}, 
    \; \theta_X > U \}\qquad (U\sim \Unif{{\mathbb U}}).
\end{align*}
These sets do, however, have the property
\[ \Pr\{ k \in \mathbf{e}_{\theta}(U) \} = \theta_k \qquad(k=1,\ldots,K), \]
which is consistent with, but weaker than, a multinomial sampling model.

For inference on $\theta$ for given $X$, combining the \dDSM (\ref{eq:ddsm}) with the
observation model $\mathbf e_{4,X}$~\eqref{eq:PthetapiX} produces
\begin{align} \label{PthetagivenX}
    \mathbf{e}_{X}(U) 
    & =  p_2( \mathbf e^*(U) \cap p_{1,2}(\mathbf e_{4,X})) \nonumber \\
    & = \{\theta:\; \theta\in\Theta, \theta_X > U\} \qquad (U\sim \Unif{{\mathbb U}}).
\end{align}
The sets $\mathbf{e}_{X}(U)$ are exactly analogous to the DS model for
a Bernouli parameter when a single ``success'' has been observed;
namely, the parameter is modeled as exceeding a $\Unif{{\mathbb U}}$ variate.

We acknowledge that the {\em a posteriori} choice to believe $\pi(1) = X$, as represented by $\mathbf e^*$~\eqref{eq:orderRestrictedDSM}, is unusual because it depends on the observed $X$.  This appears to create a paradox, which we call the \textit{restricted permutation paradox}: the permutation $\pi$ must be known to generate $X$, while simultaneously $X$ must be observed to determine $\pi$.  This issue was unresolved in an initial draft of this paper; in this draft, we provide an alternative justification for this \textit{a posterori} update that does not rely on ``re-observing" an auxiliary variable.  This discussion is found in the Supplementary Materials and this issue is discussion more broadly in Section \ref{sec:discussion} \citep{lawrence2024}.  We now extend this model to $n > 1$.

\section{Dirichlet-DSM for \texorpdfstring{$n\geq 1$}{} }
\label{sec:generaln}
To model the $n$ categorical outcomes of a multinomial draw ${X}= (X_1,\ldots,X_n$), we
combine independent sets $\mathbf e^*(U_i)$ given by~\eqref{eq:ddsm} and
restrict to non-null intersections. Specifically the \dDSM (for $n\geq
1$) is
\begin{align}
  \label{eq:fullddsm}
  \mathbf{e}_n({U}) 
  = & 
  \{({X},\theta):
    \; \theta\in\Theta,
    \ X_i\in\mathbb{X},
    \ \theta_{X_i}>U_i, 
    \ i=1,\ldots,n \}, \\
     & \left(U_i \stackrel{iid}{\sim} \Unif{0,1}
           \ \left|\; \mathbf{e}_n({U})\neq\emptyset \right.\right). \nonumber
\end{align}
where ${ U}=(U_1,\ldots,U_n)$.


The posterior model for $\theta$ is obtained by
projecting~(\ref{eq:fullddsm}) to $\Theta$ for given $ X$,
which is equivalent to independent combination of the posterior sets
$\mathbf{e}_{X_i}(U_i)$ given by~(\ref{PthetagivenX}) with restriction
to non-null intersections.
The combined posterior DSM has random sets
\begin{align} \label{PthetagivenXi}
  \mathbf{e}_{ X}({U}) 
  = & \;
  \{\theta:
  \; \theta\in\Theta,
  \ \theta_{X_i}>U_i, 
  \ i=1,\ldots,n \}, \\
  & \left(U_i \stackrel{iid}{\sim} \Unif{0,1}
           \ \left|\; \mathbf{e}_{ X}({U})\neq\emptyset \right.\right). \nonumber
\end{align}
The following theorem expresses this posterior DSM in terms
of a $K+1$ class Dirichlet distribution.

\begin{thm}
\label{thm:posterior01}
Given a sample $X_1,\ldots,X_n \stackrel{iid}{\sim} \text{Multinomial}(1, \theta_1,\ldots,\theta_K)$,
the posterior random set~(\ref{PthetagivenXi}) is distributed as
\begin{equation*}
    \mathbf{e}_{N}(Z) = \{\theta:\; \theta \in\Theta,\,
    \theta_1\geq Z_1,\ldots,\theta_K\geq Z_K\}
\end{equation*}
where
\begin{equation}
    (Z_0,Z) \equiv (Z_0,Z_1,\ldots,Z_K) \sim \mbox{Dirichlet}(1, N_1, \ldots, N_K)
\label{eq:DP08}
\end{equation}
with
\begin{equation*}
  N_k = \#\{X_i: X_i=k\}
\end{equation*}
denoting the cell counts for $k=1,\ldots,K$.  Note that $Z_k=0$ with probability one iff
$N_k=0$.
\end{thm}

\noindent
{\it Proof.}
Define
\[ Z_k = \max\{U_i:\; X_i=k, i=1,\ldots,n\} \quad (k=1,\ldots,K)\]
and $Z_0=1-(Z_1+\cdots+Z_K)$.  Then
$\mathbf{e}_{X}(U)=\mathbf{e}_{N}(Z)$ and it
remains to show that $(Z_0,Z_1,\ldots,Z_K)$ has the given Dirichlet
distribution.

Without imposing the condition
$\mathbf{e}_{X}(U)\neq\emptyset$, the CDF of
$Z_k$ is
\begin{equation*}
  \Pr(Z_k\leq z_k)
  = \Pr(U_i \leq z_k \text{ for each } i \text{ such that } X_i=k) \\
  = z_k^{N_k}
\end{equation*}
where $0\leq z_k \leq 1$ for $k=1,\ldots,K$.  Thus, the density of
$Z_k$ is proportional to $N_kz_k^{N_k-1}$ where, if
$N_k=0$, the expression is taken as the limiting Dirac delta function
for a unit mass at $z_k=0$.
From independence, the pdf of $Z_1,\ldots,Z_K$ is proportional to
$\prod_{k=1}^K N_k z_k^{N_k-1}$ and this is true both before and after
conditioning on
$\mathbf{e}_{X}(U)\neq\emptyset$\,---only the
domain of $Z$ and constant of proportionality are modified by conditioning.

The constraint $\sum_{k=1}^K \theta_k=1$ implies that
$\mathbf{e}_{X}(U)\neq\emptyset$ is
achieved by restricting $Z_1 + Z_2 + \ldots +Z_K \leq 1.$
Hence, the density of $(Z_1,\ldots,Z_K)$ is
\[ f_Z(z) \propto \prod_{k=0}^K N_k z_k^{N_k-1}
\quad \left(\sum_{k=0}^K z_k=1; z_k>0, k=0\ldots,K \right)
\]
where $z_0 \equiv 1-\sum_{k=1}^K z_k$ and $N_0\equiv 1$.
Noting that this is the density of the Dirichlet distribution
(\ref{eq:DP08}) completes the proof.  \eop

\vspace{0.1in}

\begin{figure}[th]
\centering
\includegraphics*[width=2.5in]{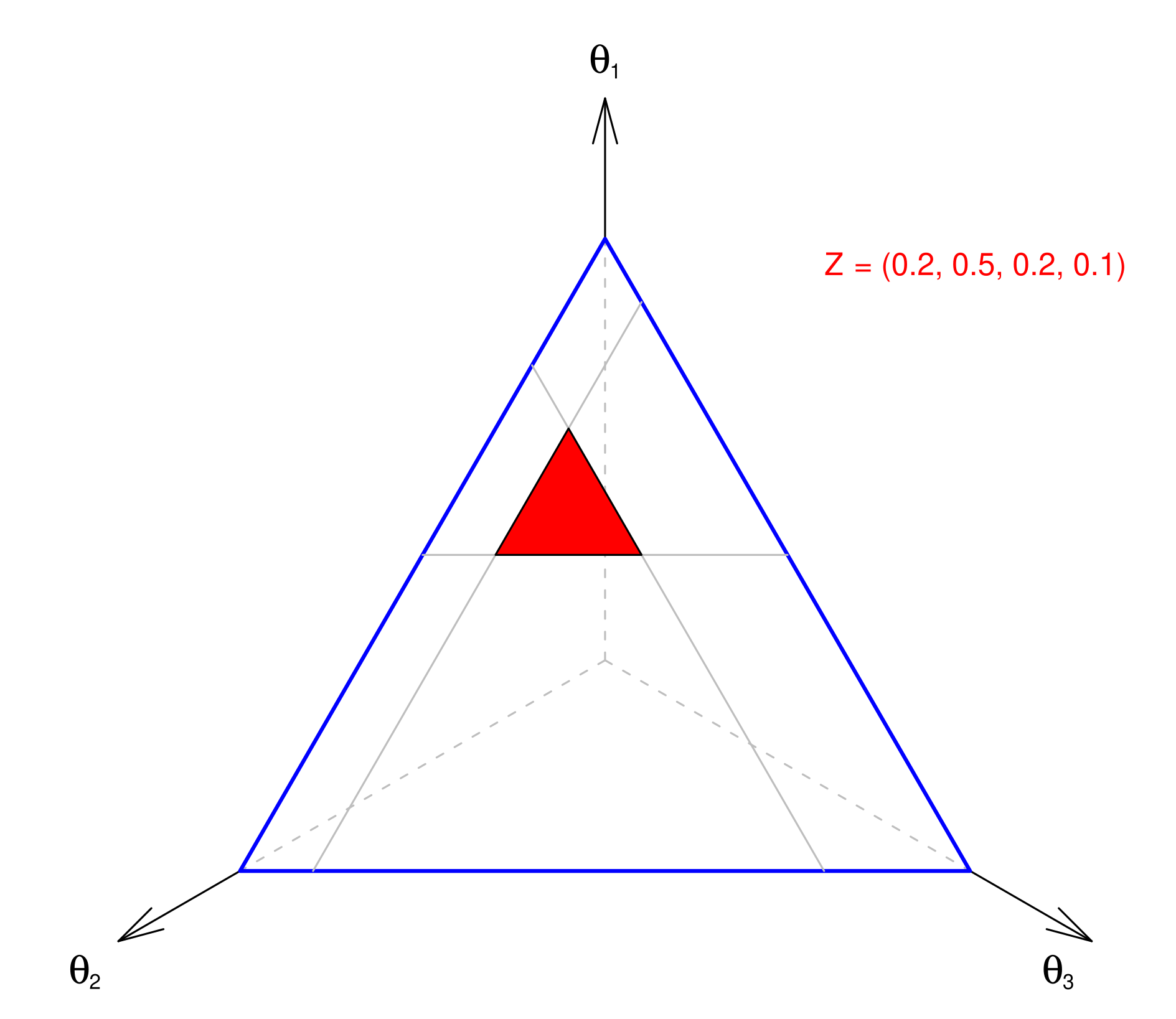}
\caption{ A realization of the random set of the posterior
distribution of $\theta$ in the trinomial model.
The parameter space $\Theta$ is shown by
the outer large triangle, representing the (standard) $3$-dimensional simplex.
The inner shaded triangle
is a draw of the random set from $\mbox{Dirichlet}(1, N_1,N_2,N_3)$ with $Z=(0.2, 0.5, 0.2, 0.1)$, with $N_k > 0$ for $k = 1, 2$, and $3$.  Note that the random set is similar to the standard simplex $\Theta$.
}
\label{mul}
\end{figure}

The random set of the posterior DS model is
a small simplex similar to $\Theta$, but positioned inside $\Theta$.
The location of this random set is determined by $Z=(Z_1,\ldots,Z_K)$.
The size of the random set is determined by $Z_0=1-(Z_1+\ldots+Z_K)$, which contributes to the $r$ component of DS inference.
Thus, it follows from (\ref{eq:DP08}) that  $r$ vanishes for large $n$.
A posterior random set in the trinomial case is depicted in Fig. \ref{mul}
as an illustrative example.

\section{Properties of  Dirichlet-DSM}
\label{sec:properties-ddsm}
The \dDSM has a number of attractive properties.  One of the most
important is practical:  its basis in the Dirichlet distribution makes
it simple to compute posterior probabilities.  This is an important
advantage over the \sDSM which has been slow to gain acceptance for practical use, in part because
its intersections are harder to compute.
In fact, the computational burden for the \dDSM is on par
with frequentist and Bayesian calculations for the Multinomial model,
but with the benefit of the improved inference that comes from
set-based probabilities.

Walley cites several properties that give IDM important advantage over traditional Bayesian and
frequentist techniques.  The \dDSM shares these principles and has an
important advantages over IDM involving inference for point
assertions and restricted models.  The \dDSM also has an important neutrality property (described in section 5.2) shared with IDM but not \sDSM.

\subsection{Walley's Principles}
\cite{walley1996} lays out three appealing inferential properties that the
imprecise Dirichlet model satisfies.  \cite{bernard2005} clarifies these
ideas and provides the descriptions used here.
\begin{description}

\item[\it The symmetry principle] Prior uncertainty should be
  invariant to permutations of the categories.  Consider
  Equation~\eqref{eq:fullddsm} that defines the \dDSM.  The focal elements
  do not depend on the ordering of the categories in any way.
  Contrast this with Equation~\eqref{eq:IXtheta} for the \iDSM where the ${\mathbb
    U}_k$ depend on partial sums of $\theta$ elements with a known,
  fixed ordering.  Since the \dDSM focal elements remain unchanged by
  permutations the symmetry property holds.  The \sDSM
  also satisfies this property.

\item[\it The embedding principle]  Prior uncertainty about any assertion
  $A$ with respect to $X$ or $\theta$ should be
    invariant to refinements and coarsenings of the
    categories so long as $A$ is fundamentally unchanged.  This is trivially true for \dDSM because
    every set $\mathbf{e}_n(U)$ in equation~(\ref{eq:fullddsm})
    projects to all of $\mathbb X^n$ and to all of $\Theta$.
    That is, \dDSM is vacuous with respect to both $X$ and $\theta$.
    Therefore, every non-trivial assertion (i.e., neither $A$ nor
    its compliment is empty) has $r=1$ regardless of any refinement or
    coarsening.

\item[\it The representation invariance principle]  This is the
  posterior version of the embedding principle for inference about
  $\theta$.
  The simplest case is when an assertion $A$ makes a claim about a single category,
  $\theta_k$.  If the other categories are pooled or
  divided in any way, uncertainty about $A$ should be unchanged.
  The same should be true if category $k$ is split in two and $A$
  is redefined by replacing $\theta_k$ with the sum of its two
  components.  This property arises from the Dirichlet distribution
  underlying \dDSM.  A formal statement and proof are
  given in the Supplementary Materials \citep{lawrence2024}.

\end{description}

Thus, the \dDSM is philosophically on par with IDM with respect to Walley's principles while maintaining other inferential advantages.

\subsection{Conditional Models for Simplex-DSM and Dirichlet-DSM}
\cite{connor1969} present the concept of {\em neutrality}.  A
vector of random proportions, $P=(P_1,\ldots,P_K)$ such that $\sum_{i} P_i = 1$,
is neutral if $P_1$ is independent of the vector $P_{-1}\equiv
(P_2,\ldots,P_K)/(1-P_1)$.
The two theorems below compare \sDSM and \dDSM with respect to
neutrality for the case $n=1$ and $X=k$. Without loss of generality, let $k=1$ for the remainder of this section.

\subsubsection{Conditional DSMs for \texorpdfstring{$\theta_k$}{} given \texorpdfstring{$\theta_{-k}$}{} and \texorpdfstring{$X=k$}{}}
Conditioning on $\theta_{-1}$~\eqref{eq:relativeTheta} implies that the relative cell probabilities in all categories excluding category 1 are known.  The question is whether $\theta_{-1}$ provides additional information on $\theta_1$.  For \dDSM it is clear from \eqref{PthetapigivenX} that knowing $\theta_{-1}$ does not affect inference on $\theta_1$ when $X=1$.  The posterior random intervals for $\theta_1$ are $[U, 1)$ where $U \sim \Unif{[0,1)}$.  This also holds for $\sDSM$ as stated in the following theorem.

\begin{thm} \label{THM:NEUTRALITY1}
Consider the $\sDSM$ for
$\mbox{Multinomial}(n, \theta_1,\ldots,\theta_K)$ with $n=1$ and $X=1$.
Given $\theta_{-1}\equiv(\theta_2,\ldots,\theta_K)/(1-\theta_1)$, the conditional DSM
for $\theta_1$ has random intervals of the form  $[U, 1)$ where $U
\sim \Unif{[0,1)}$.
\end{thm}

The proof is given in the Supplementary Materials \citep{lawrence2024}.  Thus, both DSMs have the
appealing feature that the distribution of a single element of $\theta$ is not influenced
by the known relative frequencies of the remaining elements.  We see
in the next section that the reverse is true for \dDSM but not \sDSM.

\subsubsection{Conditional DSMs for \texorpdfstring{$\theta_{-k}$}{} given \texorpdfstring{$\theta_k$}{} and \texorpdfstring{$X=k$}{}}

\begin{figure}
\begin{center}
\includegraphics*[width=4.5in]{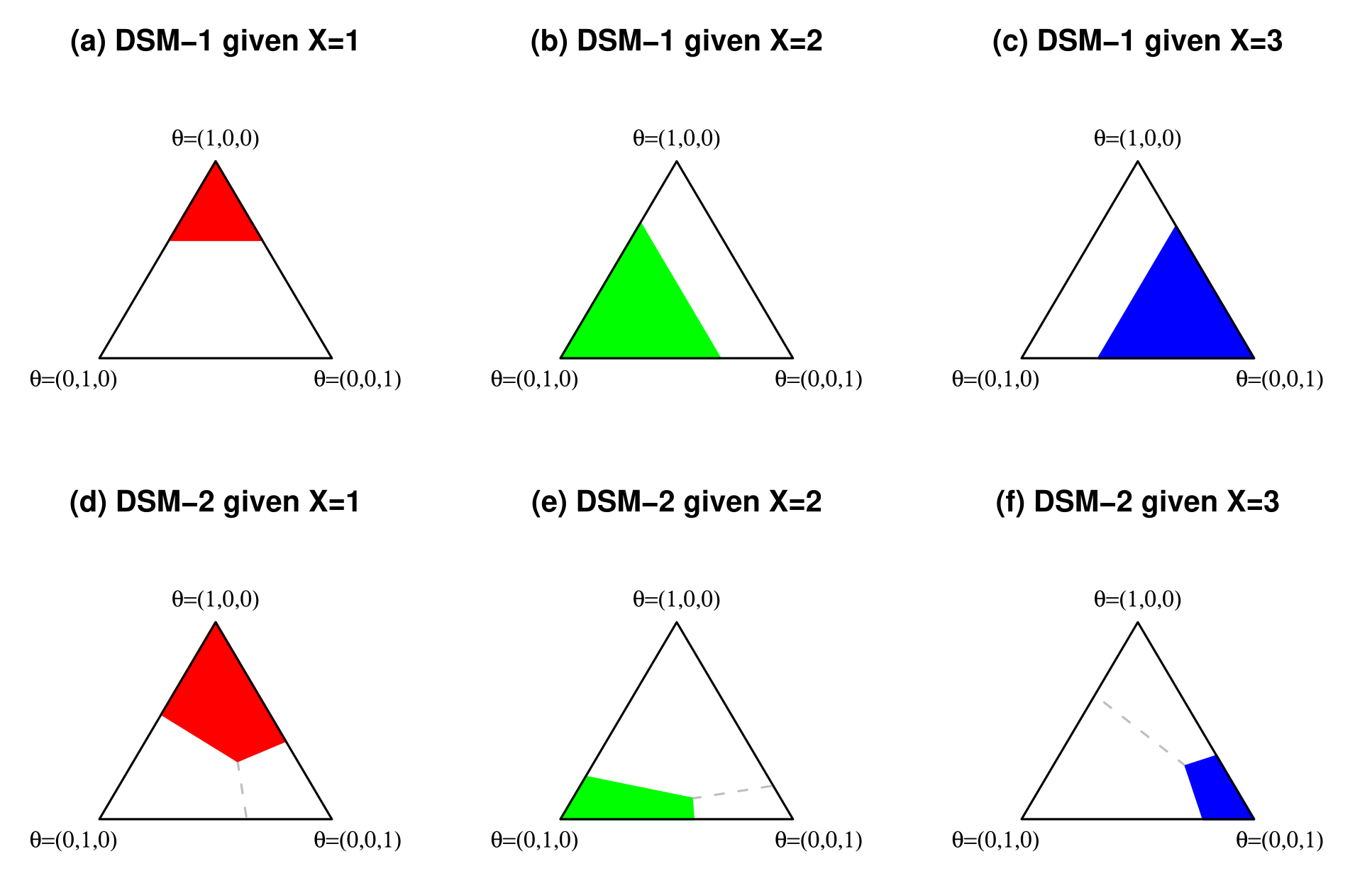}
\end{center}
\noindent
\caption{
Three examples of the focal element in the posterior $\dDSM$ [$\sDSM$] are shown
in plots (a), (b), and (c) [(d), (e), and (f)].
}
\label{fig:SDSM3-GDSM3}
\end{figure}

In the $K=3$ case with $n=1$, the focal elements in the $\dDSM$ for
$\theta$ take one of the three types displayed in the upper panel of
Fig. \ref{fig:SDSM3-GDSM3}, while the focal elements in the $\sDSM$
for $\theta$ take one of the three types displayed in the lower panel
of Fig. \ref{fig:SDSM3-GDSM3}. Given $n=1$ and $X=1$, the conditional
$\dDSM$ for $\theta_2/\theta_3$ given $\theta_1=p_1$ is obtained by
ruling out focal elements that do not contain $\theta_1=p_1$ and
intersecting the others with $\theta_1=p_1$. It follows that
the conditional DSM for $\theta_2/\theta_3$ given $\theta_1=p_1$ and
$X=1$ in $\dDSM$ is vacuous.
Similar results hold for $\dDSM$ in the general $K$ case, where the
conditional DSM places unit mass on the $K-2$ simplex, ${\cal S}_{K-2}$.  However, this
desirable property does not hold for $\sDSM$.  The following theorem
characterizes the mass of the vacuous set ${\cal S}_{K-2}$ for
$\theta_{-1}|\theta_1$ in the posterior $\sDSM$.

\begin{thm} \label{THM:NEUTRALITY2}
Consider $\sDSM$ with $n=1$ and $X=1$.  The conditional posterior
model for
$\theta_{-1}$ given $\theta_k$ gives the vacuous focal element
${\cal S}_{K-2}$ (the $K-2$ simplex) a mass of
\[
\Pr({\cal S}_{K-2}) =
    \frac{1}{1+(K-2)(1-\theta_1)}
    \frac{1}{1+\sum_{i=1}^{K-2}(1-\theta_1)^i}
\]
\end{thm}

The proof is given in the Supplementary Materials \citep{lawrence2024}.

For $\theta_1 \approx 1$, the conditional posterior of $\sDSM$ is
approximately vacuous. However, for $\theta_1 \approx 0$, the vacuous focal
element in the conditional posterior $\sDSM$ has approximate mass
$(K-1)^{-2}$ which goes to zero quickly as $K$ increases and implies
that \sDSM imposes certain information on $\theta_{-1}$ when there are
no observations to distinguish these relative cell probabilities.

We see that the \sDSM has a disturbing feature that we explicitly seek
to avoid with the \dDSM:  a single observation and the knowledge of
its probability provides information about the proportional values of
the remaining probabilities.  We simply do not believe that this
should be true, hence our decision above to ignore such information in
the \dDSM.

\section{Examples}
\label{sec:examples}

\subsection{A simple trinomial example: comparison of Bayes, IDM, Simplex-DSM, and Dirichlet-DSM}
\begin{figure}
\begin{center}
\includegraphics[width=0.65\linewidth]{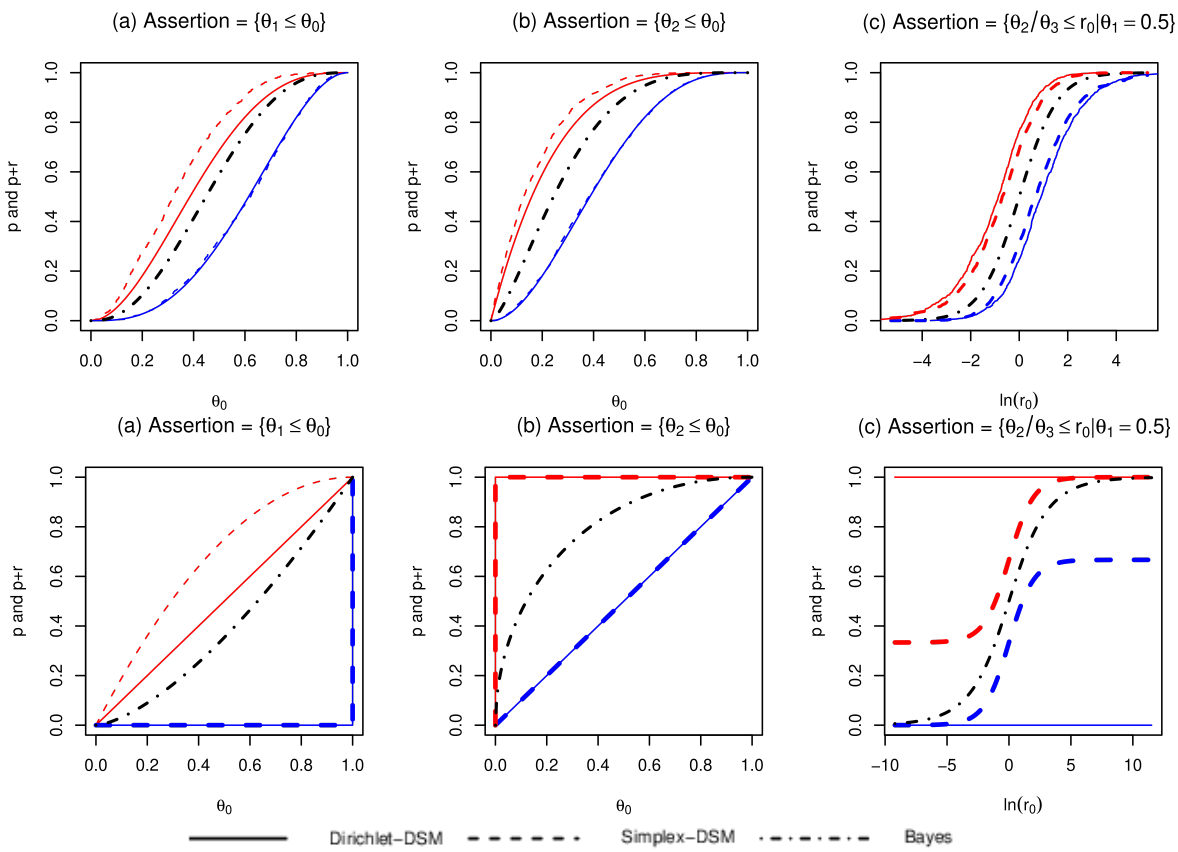}
\end{center}
\noindent
\caption{Numerical results for a trinomial with the observed
  data $N=(1,0,0)$ (top row) and $N=(2,1,1)$ (bottom row).  The
  probability $p(\theta_0)$ and plausibility $p(\theta_0)+r(\theta_0)$
  are for
(a) the assertion $\{\theta_1 \leq \theta_0\}$,
(b) the assertion $\{\theta_2 \leq \theta_0\}$ ($\{\theta_3 \leq \theta_0\}$ is identical), and
(c) the assertion $\{\theta_2/\theta_3 \leq r_0 | \theta_1 = .5 \}$
for a sequence of values of $\theta_0$ or $r_0$.
The solid lines are the $\dDSM$ results (IDM is identical), the dashed
lines are the corresponding $\sDSM$ results, the dot-dashed lines
are the Bayesian results with the Jeffreys prior.}
\label{fig:trinomial}
\end{figure}

As a simple illustrative example, consider the trinomial model $(K=3)$
with observed counts $N=(N_1, N_2, N_3)$ and unknown cell
probabilities $\theta=(\theta_1, \theta_2, \theta_3)$.  We examine two
types of assertions:  (i) $\{\theta_k\leq \theta_0\}$ with fixed
$\theta_0$ for each $k$, and (ii) $\{\theta_2/\theta_3 \leq r_0 |
\theta_1 = .5\}$ with fixed $r_0$.
Figure \ref{fig:trinomial} shows numerical results for observations
$N=(1,0,0)$ and $N=(2,1,1)$ and both types of assertions using
$\dDSM$, $\sDSM$,  Bayes, and IDM.  The
necessary calculations for the \dDSM and a summary of the
calculations for the remaining methods are available in the Supplementary Materials \citep{lawrence2024}.

The $N=(1,0,0)$ results provide a numerical illustration of the
properties of the \dDSM.  With only a single data point, we have focal
elements of the form $\{\theta: \theta_1 \geq U\}$ where is $U \sim
\Unif{[0,1)}$.  Sets of this form can never fall within the assertion
set $\{\theta_1 < \theta_0\}$, thus $p=0$ in the upper left panel of
Figure \ref{fig:trinomial}.  Sets of this form overlap the assertion
set $\{\theta_1 < \theta_0\}$ whenever $U \leq \theta_0$.  This gives
us the diagonal solid line for $p+r$ in the upper left panel.  The
assertions $\{\theta_2 < \theta_0\}$ and $\{\theta_3 < \theta_0\}$ are
converse in some sense.  Sets of the form $\{\theta: \theta_1 \geq
U\}$ always overlap these assertions and fall within them whenever $U
\geq \theta_0$.  Finally, the conditional assertion is vacuous as the
random sets always overlap, and indeed contain, this assertion for any
$r_0$.  The other methods are slightly less appealing here.  The \sDSM
has a uniformly larger value of $r$ for the first assertion and gives
a non-vacuous, and therefore unappealing, result for the conditional
assertion.  IDM matches the \dDSM for these assertions.

The $N=(2,1,1)$ results show the two DS analyses begin to approach
each other.  The \sDSM has somewhat larger values of $r$ with a
greater difference on the $\theta_1$ assertion.  In this sense, it is
more cautious than the \dDSM.  The results for the conditional
assertion are also similar, but this time the \dDSM has the larger $r$
value and has moved quite drastically from the $N=(1,0,0)$ case.

The comparison between \dDSM and \sDSM and the $N=(1,0,0)$ and
$N=(2,1,1)$ cases demonstrates the idea of `strengthening' and `weakening'
beliefs.  The \dDSM has been constructed to strengthen inference on
categories that have been observed and weaken it elsewhere.  Hence, it
provides a smaller $r$ value than \sDSM for the first assertion,
particularly with $N=(1,0,0)$.  Similarly, the ratio of the
probabilities for the unobserved categories is completely vacuous for
$N=(1,0,0)$, the `weakest' possible inference.  It is interesting that
once these categories are observed at $N=(2,1,1)$, the \dDSM inference
approaches the \sDSM inference very rapidly.  Thus, it appears that
\dDSM is a fast learner: agnostic when data are lacking, but a strong
believer thereafter.

The \dDSM and IDM analyses for this set of problems are identical, but
this is not true in general as the next example will demonstrate.

\subsection{A constrained multinomial: the linkage model}
\label{sec:real-data-example}
This example concerns the well-known linkage model of \cite{rao1973}.  The data consist of 197 animals that are distributed
into four categories, for which the theoretical population cell
probabilities are
\begin{equation}
\theta = \left(\frac{1}{2}+\frac{\phi}{4},
\frac{1-\phi}{4},\frac{1-\phi}{4},\frac{\phi}{4}\right)
\label{eq:linkage01}
\end{equation}
for some
$\phi$ with $0<\phi<1.$ The observations are
$N=(125,18,20,34)$ but in order to more clearly evaluate the
effects of modeling, we consider counts reduced by a factor of
approximately five:  $N=(25, 3, 4, 7)$.  Results using the original
counts are similar, but model differences are easier to discern with
the reduced data.  For DS analysis, the inference about $\phi$ is
obtained directly from the DSM for the {\it saturated} ($4$-cell)
multinomial model by combining it with assumption
(\ref{eq:linkage01}), regarded as a DSM.

The random region for $\theta=(\theta_1,\theta_2, \theta_3,\theta_4)$ with the constraint that $\pi(1) = X$ in the saturated multinomial model
is given by
\[
       \theta_1 \geq Z_1,\,
       \theta_2 \geq Z_2,\,
       \theta_3 \geq Z_3,\, \mbox{ and }\,
       \theta_4 \geq Z_4,
\]
where
\( (Z_0,Z_1,\ldots,Z_4)\sim \mbox{Dirichlet}(1,25, 3, 4, 7)\) for the
$\dDSM$ (Theorem \ref{thm:posterior01}).  Thus, given that $\theta$ is
of the form (\ref{eq:linkage01}), the conditional/combined posterior
for $\phi$ has a random set of the form
\begin{equation}
      \phi \in \left[ \max(4Z_1-2,4Z_4),\; \min(1-4Z_2,1-4Z_3) \right]
\label{eq:a-interval}
\end{equation}
which is obtained by ruling out the {\it conflict} cases
(Dempster 2008):
\[
       \max(4Z_1-2,4Z_4)> \min(1-4Z_2,1-4Z_3).
\]

The IDM analysis is described in detail in the Supplementary Materials \citep{lawrence2024}.  Essentially, the standard IDM parameters are transformed to $\phi$ and two nuisance parameters are conditioned away.  The resulting IDM family of priors for $\phi$ is no longer vacuous and not symmetric.

\begin{figure}[t]
\begin{center}
\includegraphics*[width=5in]{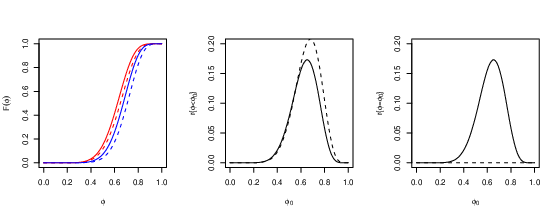}
\end{center}
\label{fig:linkage}
\caption{Numerical results for the linkage example.  The left panel
  shows the upper (red) and lower (blue) CDFs for the \dDSM (solid)
  and IDM (dashed).  The remaining panels show $r$ values for $\{\phi
  \leq \phi_0\}$ (middle) and for $\{\phi = \phi_0\}$ (right) with a
  solid line \dDSM and a dashed line for IDM.  The \dDSM r values are
  identical in the middle and right plots.}
\end{figure}

The left two panels of Fig. \ref{fig:linkage} summarize the results
for \dDSM (solid lines) and IDM (dashed lines) on assertions of the
form $\{\phi \leq \phi_0\}$.  The first panel gives the upper (red)
and lower (blue) CDFs for these models.  IDM favors larger
values of $\phi$.  An examination of the priors reveals that the upper
bounding prior is proportional to $[(2+\phi)(\phi)(1-\phi)]^{-1}$.
This is approximately bathtub-shaped with more than half of the mass
shifted toward zero, but roughly symmetric.  The posterior for this
case appears in the middle of the DS CDFs.  The lower bounding prior
is proportional to $[(2+\phi)(1-\phi)(1-\phi)]^{-1}$, a distribution
with the vast majority the mass near one.  This example illustrates an
important problem with IDM: it is not designed for restricted
parametrizations.  The family of priors that is induced by a
restriction may be informative in unexpected ways.  A
second apparent result of this
issue is shown in the second panel of Fig. \ref{fig:linkage}.  The IDM
has a larger value for $r$.

The right panel in Fig. \ref{fig:linkage} summarizes the results for
\dDSM (solid red line) and IDM (blue dashed line) for assertions of
the form $\{\phi = \phi_0\}$.  The figure plots $r$ across the range
of $\phi_0$.  For both analyses, $p=0$ uniformly.  For the IDM, we also have $r=0$ uniformly.
The \dDSM, on the other hand, allows the plausibility of this type of
assertion.
In fact, the $r$ curve is identical to the one for $\{\phi
\leq \phi_0\}$ because a random focal element that contains the single
point given in the assertion will also contain neighborhoods just
below and just above the point.
This reinforces the claim that \dDSM gives sensible results over a wider range
of assertions.  This same experiment was repeated in \cite{jacob2021} comparing the \dDSM against the \sDSM; the results in this paper were also comparable.

\section{Computational Advantages over the \sDSM} \label{sec:computational_advantages}

As visualized in Figure \ref{fig:SDSM3-GDSM3}, taking samples from a \sDSM involves distributions of convex polytopes within the simplex.  There is currently no closed-form distribution for sampling the vertices of these polytopes, which made the original formulation of the \sDSM computationally impractical.  This issue has been addressed directly in \cite{jacob2021}, where a Gibbs Sampler is developed that converges to a uniform distribution on the \sDSM conditioned on the observed data.

While recent theory certainly improves upon the feasibility of the \sDSM, and offers a fascinating connection between simplex-based DS sampling and graph theory, the \sDSM still falls well short of the computational simplicity of the \dDSM.  In the following, we will review the two applications considered in \cite{jacob2021} and compare the runtimes between the two methods.

\subsection{The Linkage Model}
We return to the linkage model example from Section \ref{sec:real-data-example} as a means to compare the runtimes between the \sDSM and the \dDSM.  For the \sDSM approach, we run the Gibbs sampler from \cite{jacob2021} first on the $K = 4$ Multinomial model ignorant of the linear constraint in \eqref{eq:linkage01}.  Then, of the feasible sets generated, only those that obey \eqref{eq:linkage01} are kept\footnote{This is identical to the approach proposed in \cite{jacob2021}.}.

\begin{table}
\centering
\begin{tabular}{ |p{3cm}||p{1.6cm}|p{1.6cm}|p{3cm}|p{3cm}|  }
 \hline
 \multicolumn{5}{|c|}{Runtime (in seconds) and Acceptance Rates of DS Approaches on the Linkage Model} \\
 \hline
 \makecell{Model} & \makecell{Mean\\ Runtime}  & \makecell{St.Dev.\\ Runtime} & \makecell{Mean Rate\\ Acceptance} & \makecell{St.Dev. Rate\\ Acceptance} \\
 \hline
 \sDSM & 33.36 & 0.592 & 0.051 & 0.001  \\
 \dDSM & 0.063 & 0.012 & 0.042 & 0.001 \\
 \hline
\end{tabular} \caption{Metrics on runtime and acceptance rates when fitting the \sDSM and \dDSM.  Mean and Standard Deviation are calculated by repeating the experiment 30 times on a 12-core Windows computer. 
A single experiment of the \dDSM takes 112500 samples from a Dirichlet distribution; a single experiment of the \sDSM starts 250 separate chains that run for 500 iterations, throwing out the first 50 as burn-in.  Each of these processes produce 112500 samples from the post-data DSM.
} \label{tab:runtime_linkage}
\end{table}

We compare the runtimes and acceptance rates between the two approaches in Table \ref{tab:runtime_linkage}. The \dDSM is dramatically faster than the simplex sampling method for the \sDSM.  The acceptance rates are about the same, with the \sDSM fit accepting $\phi$ bounds that obey \eqref{eq:linkage01} slightly more frequently than the \dDSM fit.  The authors in \cite{jacob2021} point out that while these acceptance rates are low, they are feasible for practical applications.  

We examine the acceptance rates for both models for an experiment with data that conflict slightly with the constraint imposed by \eqref{eq:linkage01} as a means to see whether either model is more robust to data that appear to not obey the constraint.  Using the updated count data $N=(25,6,2,7)$, the average acceptance rates of the \sDSM and \dDSM dropped to 0.0194 and 0.0160, respectively.  Clearly, both models have comparable acceptance rates that drop whenever the data are less in-line with the linear constraint in \eqref{eq:linkage01}.  We maintain that even though both acceptance rates drop, the \dDSM remains more practical since a much larger number of samples can be taken in a reasonable amount of time, even if only a small fraction are accepted.  As suggested by \cite{jacob2021}, fitting these models on data even more in conflict with the constraint would require novel computational methods. 

\subsection{Testing Independence} \label{sec:testing_indep}

The authors of \cite{hoffman2021} make several comments comparing the runtime required to fit the \sDSM against that required to fit the \dDSM when testing for independence of two random variables on the unit cube using a multinomial model.  This somewhat generalizes the experiment considered in \cite{jacob2021}, where the hypothesis test $H_0: \theta_1\theta_4 = \theta_2 \theta_3$ is assessed for the $K = 4$ multinomial model.  We will briefly describe the experiment in \cite{hoffman2021} and expand on their results here.  Suppose $(X,Y)$ follow some distribution on $[0,1]^2$ and the aim is to determine whether $X \perp\kern-5pt\perp Y$.  To do so, consider a discretization of the sample space $\Omega = [0,1]^2$,
\[ \Omega = \cup_{i=1}^\mathcal{K} \cup_{j = 1}^\mathcal{K} I_i^\mathcal{K} \times I_j^\mathcal{K}, \]
where $I_i^\mathcal{K} = \left[ \frac{i-1}{\mathcal{K}}, \frac{i}{\mathcal{K}} \right)$.  These intervals determine the count statistics,
\[ Z_{i,j} = \# \{x : (X_x, Y_x) \in I_i^\mathcal{K} \times I_j^\mathcal{K} \}, \]
\sloppy for $x \in \{1, \dots, n\}$, which create a $\mathcal{K}\times \mathcal{K}$ contingency table.  Given $n$ observations $(X_1, Y_1), \dots, (X_n, Y_n)$, we use a Multinomial distribution to model these data organized into a contingency table,
\[ (Z_{11}~\dots~Z_{\mathcal{K}\mathcal{K}} | \sum_{i,j \in \{1,\dots, \mathcal{K}\}} Z_{i,j}) \sim \text{Multinomial}(n, (\theta_{11}, \dots, \theta_{\mathcal{K}\mathcal{K}})). \]
For $\mathcal{K} = 2$, this problem is equivalent to the $K = 4$ Multinomial model ($K = \mathcal{K}^2$ generally). considered in \cite{jacob2021}.  To determine independence, several samples of $(\theta_{11}, \dots, \theta_{\mathcal{K}\mathcal{K}})$ are taken via either the \dDSM or the \sDSM.  If these samples are sufficiently far from $(1/K, \dots, 1/K) \in [0,1]^K$, then the assumption of independence is rejected.  This experiment is more general than the one considered in \cite{jacob2021} since the granularity of the discretization $\mathcal{K}$ can be increased and the two models can be compared on increasingly complex simplexes.  

\begin{table}
\centering
\begin{tabular}{ |p{3cm} p{2cm} |c|c|  }
 \hline
 \multicolumn{4}{|c|}{Runtime (in seconds) DS Approaches for Testing Independence} \\
 \hline
 \multicolumn{2}{|c|}{Model} & \makecell{Mean\\ Runtime}  & \makecell{St.Dev.\\ Runtime} \\
 \hline
  \sDSM & $2 \times 2$ & 10.493 & 1.777 \\
  ~ & $3 \times 3 $& 61.980 & 16.647 \\
  ~ & $6 \times 6$ & $> 2$ days & NA \\
  ~ & $8 \times 8$ & $> 2$ days & NA \\
 \dDSM & $2 \times 2$ & 0.928 & 0.199 \\
  ~ & $3 \times 3$ & 1.040 & 0.231 \\
  ~ & $6 \times 6$ & 1.559 & 0.315 \\
  ~ & $8 \times 8$ & 2.036 & 0.506 \\
 \hline
\end{tabular} \caption{Metrics on runtime when fitting the \sDSM and \dDSM for the test of independence described in Section \ref{sec:testing_indep}.  Mean and Standard Deviation are calculated by repeating the experiment 30 times given a single core on a Linux HPC environment.  The Gibbs sampler for \sDSM was run for 40 steps, with a 10-step burn-in. } \label{tab:runtime_indep}
\end{table}

We modify and expand upon the original experiment in to compare the runtimes of the \sDSM and the \dDSM.  For each model, we fit the model under the null hypothesis and perform the test for independence.  We ran the Gibbs sampler used to fit the \sDSM for a relatively short number of iterations (40 total with 10 burn-in); the \dDSM only requires a single iteration to run since it is functionally equivalent to taking samples from a Dirichlet distribution.  This entire experiment was run on an Linux HPC environment and each model was given a single core.  It is generally advisable to give the Gibbs sampler several cores to run multiple chains simultaneously, but in this small-scale experiment we found it appropriate to give each model the same computational resources.  Each simulation was run 30 times and the mean and standard deviations of the resulting runtimes are given in Table \ref{tab:runtime_indep}.  The \dDSM model is clearly the faster model, with the \sDSM not completing every experiment in the 2 days it was given to run. 

The authors of \cite{hoffman2021} take this experiment a step further and perform a simulation study on the ability of each model to correctly reject an assumption of independence; this includes an assessment of the power of each method at higher and higher granularities.  These authors found mostly identical performance between the two methods, with the \dDSM having more power to reject the null hypothesis.  This, along with the fact that only the \dDSM can handle granularities above $\mathcal{K} = 3$, suggest that the \dDSM is the superior method for this problem.

\section{Conclusion}\label{sec:discussion}

The two conventional DS models, the $\iDSM$ and \cite{dempster1966}'s
$\sDSM$, for analysis of multinomial data are briefly reviewed to
demonstrate the need for a better DSM for multinomial inference.  The
$\iDSM$ is attractive because the fundamental idea behind DS of
``continuing to believe'' is applied to predicting a single univariate
auxiliary random variable per observation rather than a $K-1$
dimensional auxiliary variable, as in the $\sDSM$.  However, the \iDSM
posterior is undesirable because it depends on the ordering of
categories. We also find that $\sDSM$ has undesirable properties
for inference on unobserved categories, and significant computational disadvantages.

Criticisms of Bayesian analysis for multinomial data have appeared in
different places (\cite{walley1996} and \cite{bernard2005}) and it is not our
intention to enter into this contentious argument.  The IDM approach
offers an attractive generalization of Bayesian methods that considers
a class of priors instead of a single one.  Although it provides upper
and lower probabilities, it does not contain the concept of a ``don't
know'' category \citep{dempster2008} {\it per se} as obtained from using
DS random sets.  Further, we have seen that the random set approach
produces non-trivial results over a larger class of assertions.

The \dDSM is developed by modifying the \iDSM in two ways.  First, we
consider an unknown permutation of the category intervals comprising the unit
interval.  This removes the dependence on the ordering while retaining
the attractive mapping from the uniform variable to the categories.
Second, we modify our belief about this ordering after observing the data
to avoid influencing inference on unobserved categories and to
guarantee sensible learning.  This amounts to ``strengthening" our
belief about the observed category and ``weakening'' our belief about
the unobserved categories.  An important side-benefit is that the
modified belief is more symmetric.  As shown in Section
\ref{sec:properties-ddsm}, the \dDSM achieves many, if not all, of the
desirable properties of an IDM analysis because of their shared
distributional basis.  Further, the Dempster-Shafer set-based
probability and rules of combination provide the \dDSM with
flexibility in the kinds of assertions (e.g., point assertions) and
models (e.g., constrained multinomials) that can be
considered.  Thus, it improves on the inference of both of these very
strong methods.  Finally, the \dDSM is much more computationally efficient than the \sDSM, and this aspect of traditional DS approaches is no longer a hindrance for practical use. 

Although \dDSM has several
desirable properties, the
{\em a posteriori} choice for the unknown permutation is somewhat
{\em ad hoc},
despite the fact that it chosen for the principled reasons listed.
Although we strongly feel that the \dDSM provides principled inference
for multinomial data, it is difficult to generalize the {\em ad hoc}
modification of a belief represented by eliminating undesirable
permutations from the random set (\ref{eq:PXthetapi}) to arrive at the
\dDSM model (\ref{eq:ddsm}).
As we outline in the Supplementary Materials, we believe that the \dDSM represents a weakening of the original DS argument \citep{lawrence2024}; investigating this concept is a direction for future work.

%

\begin{supplement}
\stitle{Supplementary Materials for ``A New Method for Multinomial Inference using Dempster-Shafer Theory"}
\sdescription{A further discussion on the modification to the permutation $\pi$, the proofs for two theorems in this paper, and further details on the trinomial calculation and the linkage model.}
\end{supplement}


\bibliographystyle{imsart-nameyear} 
\bibliography{bibliography}       

\newpage

\begin{center}
\huge{Supplementary Materials for ``A New Method for Multinomial Inference using Dempster-Shafer Theory"}
\end{center}

\bigskip
\section{The restricted permutation paradox} \label{subsec:paradox}
As mentioned previously (Section~\ref{sec:restr-perm}), updating the permutation $\pi$ used to generate a data point $X$ \textit{after observing $X$}, is an {\em ad hoc} constraint on $\pi$.  This represents a theoretically dubious procedure, since the generative model for $\mathbf{X} = \{X_1, \dots, X_n\}$ depends on $\pi$, and conversely $\mathbf{X}$ is needed to set the \dDSM constraint on $\pi$ (this is refered to as the \textit{restricted permutation paradox} in the main text).  In this section, we discuss another way of framing the \dDSM without introducing the unobserved auxiliary variable $\pi$ as part of the original DSM.  Like the version of the \dDSM discussed in the paper, the model proposed here is still not a DSM in the usual sense.

%
Recall the DSM representing the ``continuing to believe" step \eqref{eq:PXthetapiU} and the DSM representing the \textit{original} auxiliary equation \ref{eq:first-kind},
\begin{align} 
   \mathbf e_1(U) =& \{(X, \theta,\pi,U^*):\; X\in \mathbb{X}, \theta\in\Theta, \pi \in \Pi, U^*=U \} \qquad
   (U \sim \Unif{{\mathbb U}}), \\
\mathbf e_2 =& \{(X, \theta,\pi,U^*):\; X\in \mathbb{X}, \theta\in\Theta, \pi \in \Pi, U^* \in \mathbb{U}, X = a (\theta, U^*) \}.
\end{align}

As covered in the main text, this model corresponds to the \iDSM:

\paragraph{1) Known $\theta$.}
Knowledge of $\theta$ is represented by the DSM with
a single unit-mass focal element projected to $\mathbb{X} \times \Theta$
\[
   \mathbf e_{3,\theta} 
   = p_{1,2}(\{(X, \theta, U):
   \; X\in\mathbb{X}, 
   \; U \in \mathbb{U} \}).
\]
Combining this with (\ref{eq:IXtheta}) and projecting to $\mathbb{X}$ gives
\begin{align*}
   \mathbf{e}_\theta(U)
   & = p_1(\mathbf e_{3,\theta}  \cap \mathbf e(U) ) \\
   & =
   \{X:\; X\in \mathbb{X},
   \; X = a(\theta, U)
   \}\qquad (U\sim \Unif{{\mathbb U}})
\    
\end{align*}
which is equivalent to (\ref{eq:first-kind}), the multinomial sampling model.

Instead of incorporating an unknown permutation $\pi$ into the auxiliary equation, we instead frame the problem as a non-direct form of sampling on the parameter space.

\paragraph{2) Observe $X$.}  This corresponds to the DSM
\[
\mathbf e_{4,X} = p_{1,2}(\{(X, \theta, U):
    \; \theta\in\Theta, 
    \; U \in \mathbb{U} \}) .
\]
Combining this with (\ref{eq:IXtheta}) and projecting to $\Theta$
gives
\begin{align}
 \mathbf e_X(U) 
    & = p_2(\mathbf e_{4,X}  \cap \mathbf e(U) ) \nonumber \\
    &= \{\theta:\; \theta \in \Theta, X = a(\theta, U)\} \nonumber \\
    &= \{ \theta:\; \theta \in \Theta, 
          \theta_1 + \cdots + \theta_{X-1} \leq U < \theta_1 + \cdots + \theta_X
        \} 
        \label{eq:interval_obsX}
\end{align}
where $U\sim \Unif{{\mathbb U}}$.  

In the ``continuing to believe" step, we assume that a non-observed quantity for $U$ has been realized, and use the distribution on $U$ to generate a distribution on the parameter vector $\theta$.  The equates to sampling on a simplex -- a difficult challenge -- that the methodology in this paper circumvented by assuming that the data observed is in the first category.  The main text justifies this assumption using an unobserved permutation.  This solution has desirable theoretical properties and strong practical justifications, but results in the restricted permutation paradox.  In the following, we modify and extend an idea from \cite{hoffman2024} to resolve the restricted permutation paradox.  However, it remains true that this procedure does not result in a DSM in the usual sense.

After observing the data $X$, the ``continuing to believe" step above assumes that $U \sim \text{Unif}(0,1)$ with the set \eqref{eq:interval_obsX}.  \textit{After} the data are observed, let $\pi$ be a fixed permutation that puts category $X$ in the first position: $\pi(X) = 1$.  Define the map $Q_{\pi, X, \theta}$ as a function that takes every point on $[0,1)$, permutes the intervals determined by $(\theta, \pi)$, and maps these back onto $[0,1)$ so that 

\[ U \in \left[ \sum_{i=1}^{X-1} \theta_i, \sum_{i=1}^{X} \theta_i \right) \iff Q_{\pi, X, \theta} (U) \in  \mathbb{U}_{X, \pi}, \]
where $\mathbb{U}_{X, \pi}$ is defined in \eqref{eq:Upi} in the main text

Using this map, define the second auxiliary variable,
\[ W = Q_{\pi, X, \theta} (U). \]
In this scenario, $\pi$ and $X$ are fixed, and the map $Q$ is a bijection \textit{for every possible $\theta$}.  Thus, $W \sim \text{Unif}(0,1)$.  Note the condition that random uniform auxiliary variable $U$ to be consistent with the set \eqref{eq:interval_obsX} as implied by the fixed data observation $X$ is an equivalent condition to
\begin{equation} \label{eq:cond2}
W \leq \theta_{X},
\end{equation}
since $Q$ is trivially invertible.  Thus, sampling from the \iDSM is functionally equivalent to using the ``continuing to believe" concept on the set defined by \eqref{PthetagivenX} in the main text.  

The above is not a typical DSM because the auxiliary equation by which the data are sampled is not the same as the one used for inference on the parameter space.  We believe that this argument represents a weakening of the fiducial argument used in Dempster-Shafer Calculus.  Further investigation of this concept is a hopeful direction for future work.

\section{Proof of Theorem \ref{THM:NEUTRALITY1}}
\label{sec:neutralityOne}
The \sDSM can be described using exponential random variables for technical simplicity.  The focal elements of the \sDSM for $\theta=(\theta_1,\ldots,\theta_K)$ have the form
\begin{equation}
\frac{Z_1}{\theta_1} \leq \frac{Z_j}{\theta_j}\qquad (j=2,\ldots,K)
\label{eq:SDSM010}
\end{equation}
where $Z_i\stackrel{iid}{\sim}\mbox{Expo}(1)$ for $i=1,\ldots,K$.  Consider the conditional DSM for $\theta_1/(1-\theta_1)$ given $(\theta_2,\ldots,\theta_K)/(1-\theta_1)$.  It follows from (\ref{eq:SDSM010}) that the lower end of the random interval
for $\theta_1/(1-\theta_1)$ is given by
\[
     A = \frac{Z_1}{1-\theta_1}\frac{1}{\min_{j=2,\ldots,K} \frac{Z_j}{\theta_j}}
\]

The {\it commonality} (Shafer, 1976; Dempster, 2008) of the interval $[a, \infty]$ $(a\geq 0)$ is
\[
\Prob{A \leq a} = \Prob{(1-\theta_1) \min_{j=2,\ldots,K} \frac{Z_j}{\theta_j}
\geq \frac{Z_1}{a}}= \Prob{Z_0 \geq \frac{Z_1}{a}}
\]
where conditional on \((\theta_2,\ldots,\theta_K)/(1-\theta_1)\), $Z_0 = (1-\theta_1) \min_{j=2,\ldots,K} (Z_j/\theta_j)$ is independent of $Z_1$ and follows $\mbox{Expo}(1).$ It follows that
\[
\Prob{ A \leq a} = \frac{a}{a+1},\qquad   (a \geq 0).
\]
Let $B$ be the lower bound for $\theta_1.$ Then $A = B/(1-B)$ and
\[
\Prob{B \leq b} = \Prob{A \leq \frac{b}{1-b}}
  = \frac{b/(1-b)}{b/(1-b) + 1} = b \qquad (0\leq 0\leq 1)
\]
In other words, conditional on $(\theta_2,\ldots,\theta_K)/(1-\theta_1),$ the random interval for $\theta_1$ has the form of $[U, 1]$ with the random variable $U ~ \Unif{0,1},$ which is the same as $\dDSM$ for $\theta_1$ given $n=1$ and $X=1$ and is independent of $(\theta_2,\ldots,\theta_K)/(1-\theta_1)$.
\eop

\section{Proof of Theorem \ref{THM:NEUTRALITY2}}
\label{sec:neutralityTwo}
Without loss of generality, we take $X=1$.  The random set for
$(\theta_1,\dots,\theta_K)$ is
\[
\frac{Z_1}{\theta_1}\leq \frac{Z_j}{\theta_j} \qquad (j=2,\ldots,K),
\]
or equivalently
\begin{equation}
 \frac{\theta_j^*Z_1}{Z_j}\leq \frac{\theta_1}{1-\theta_1} \qquad (j=2,\ldots,K),
\end{equation}
where $Z_i \stackrel{iid}{\sim} \mbox{Expo}(1)$ for $i=1,\ldots,K$ and $\theta_j^*=\theta_j/(1-\theta_1)$ for $j=2,\ldots,K$.  Note that $(\theta_2^*,\ldots,\theta_K^*)\in {\cal S}_{K-2}$.  It follows that the commonality of ${\cal S}_{K-2}$ representing the vacuous information in the conditional DSM for $(\theta_2,\ldots,\theta_K)/(1-\theta_1)$ is given by the probability event that
\[
 \frac{Z_1}{Z_j}\leq \frac{p_1}{1-p_1}
\]
for all $j=2,\ldots,K$ and all $(\theta_2^*,\ldots,\theta_K^*)\in {\cal S}_{K-2}$, that is,
\[
 \max_{j=2,\ldots,K} \frac{Z_1}{Z_j} \leq \frac{p_1}{1-p_1}.
\]
Routine algebraic operations lead to
\begin{eqnarray*}
\Prob{\max_{j=2,\ldots,K} \frac{Z_1}{Z_j} \leq  \frac{p_1}{1-p_1}}
 &=& \Prob{\min_{j=2,\ldots,K}Z_j\geq \frac{1-p_1}{p_1}Z_1}\\
 &=& \int_0^\infty e^{-\frac{(K-1)(1-p_1)}{p_1}z} e^{-z}dz\\
 &=& \frac{p_1}{(K-1)(1-p_1)+p_1}\\
 &=& \frac{p_1}{1+(K-2)(1-p_1)}.
\end{eqnarray*}
This probability needs to be renormalized due to the fact that the random set in the posterior $\sDSM$ for $(\theta_1,\ldots,\theta_K)$ needs to have a non-empty intersection with the subset $\{(\theta_1,\ldots,\theta_K):\; \theta_1=p_1\}\subseteq {\cal S}_{K-1}\equiv\Theta$, which is equivalent to
\[
      U_1\equiv \frac{Z_1}{\sum_{i=1}^K Z_i} \leq p_1.
\]
The distribution of $U_1$ is given by $\mbox{pBeta}(p_1,1,K-1)$, where $\mbox{pBeta}(.,\alpha, \beta)$ denotes the the CDF of the Beta distribution $\mbox{Beta}(\alpha, \beta)$ with shape parameters $\alpha$ and $\beta$.
Thus,
\[
    \Prob{U_1\leq 1} = 1-(1-p_1)^{K-1}
           = p_1\left[1+\sum_{i=1}^{K-2}(1-p_1)^i\right].
\]
As a result, the normalized commonality of ${\cal S}_{K-2}$ is
\[
    \frac{1}{1+(K-2)(1-p_1)}\cdot \frac{1}{1+\sum_{i=1}^{K-2}(1-p_1)^i}.
\]
The theorem is proved.
\eop

\section{Representation Invariance}
\label{sec:RIP}
The definition of representation invariance for a multinomial
Dempster-Schafer model requires notation for aggregation of categories
as follows.

As in the body, let $X=(X_1,\ldots,X_n)$ represent a $K$-class
multinomial sample with corresponding cell counts $N=(N_1,\ldots,N_k)$
and probabilities $\theta=(\theta_1,\ldots,\theta_K)$.  Let
$P=\{i_1,\ldots,i_L\}$ be an arbitrary partition of
$\{1,\ldots,K\}$ and define $\tilde\theta$ as the aggregate
of $\theta$ with respect to $P$.  Namely,
$\tilde\theta=(\tilde\theta_1,\ldots,\tilde\theta_L)$ where
\[ \tilde\theta_\ell = \sum_{k\in i_\ell} \theta_k \]
for $\ell=1,\ldots,L$.
Define aggregate cell counts $\tilde N = (\tilde N_1,\ldots,\tilde N_L)$ similarly.

Consider any assertion $\tilde A$ on the space of $\tilde\theta$ and
its corresponding disaggregated version
\[ A = \{\theta: \tilde\theta \in \tilde A\}. \]
Representation invariance for a class of multinomial DSMs means that
for any partition $P$ and aggregate assertion $\tilde A$, the $(p,q,r)$
output from the DSM based on aggregate data $\tilde N$ equals the
$(p,q,r)$ output for the disaggregated assertion $A$ based on data $N$.

\begin{thm}
\label{thm:RIP}
The \dDSM is representation invariant.
\end{thm}

\noindent
{\it Proof.}
The DSM~(\ref{PthetagivenXi})for $\theta$ given $X$ is
\begin{align}
  \label{eq:ddsm1}
    \mathbf{e}_{N}(Z) =& \{\theta \in\Theta:\
                           \theta_1\geq Z_1,\ldots,\theta_K\geq Z_K\}
    \\
    & \left((Z_0,Z) \sim \mbox{Dirichlet}(1, N)\right). \nonumber
\end{align}
Define $\tilde Z = (\tilde Z_1,\ldots,\tilde Z_L)$ where
$\tilde Z_\ell = \sum_{k\in i_\ell}Z_k$ with respect to a partition $P=\{i_1,\ldots,i_L\}$.
The projection of $\mathbf{e}_{N}(Z)$ onto the space of $\tilde\theta$
is
\begin{align}
  \label{eq:ddsm2}
    \tilde{\mathbf{e}}_{N}(Z)
       &= \{\tilde\theta: \theta\in\Theta, \theta_1\geq Z_1,\ldots,\theta_K\geq Z_K\}
       \\
       &= \{\tilde\theta: \theta\in\Theta, \tilde\theta_1\geq \tilde Z_1,
                                    \ldots,\tilde\theta_L\geq \tilde Z_L\}
       \nonumber
\end{align}
The second equality can be seen by noting that every $\theta$ in the
definition of the first set also meets the aggregate inequalities for
the second set.  Furthermore, for every $\tilde\theta$ in the second
set there is {\it some} corresponding $\theta$ that meets the
disaggregated inequalities for the first set.

Clearly $\tilde{\mathbf{e}}_{N}(Z)$ depends on $Z$ only through $\tilde
Z$.  A basic property of the Dirichlet distribution is that aggregation
over a partition produces a lower dimensional Dirichlet.  Therefore,
\[ (Z_0,Z) \sim \mbox{Dirichlet}(1, N)
\quad \Rightarrow \quad
    (Z_0,\tilde Z) \sim \mbox{Dirichlet}(1, \tilde N) \]
which demonstrates that the projected model~(\ref{eq:ddsm2}) is
identical to the \dDSM for $\tilde\theta$ based directly on data
$\tilde N$.

Finally, for any aggregate assertion $\tilde A$, the events
\[ \tilde{\mathbf{e}}_{N}(Z) \subset \tilde A
   \quad\text{and}\quad
   \mathbf{e}_{N}(Z) \subset A \]
are equivalent, as are the events
\[ \tilde{\mathbf{e}}_{N}(Z) \subset \tilde A^c
   \quad\text{and}\quad
   \mathbf{e}_{N}(Z) \subset A^c \]
where $A^c$ and $\tilde A^c$ denote the complements of $A$ and $\tilde
A$ respectively.
Equivalence can be seen by noting that the events on
$\mathbf{e}_{N}(Z)$ imply those on $\mathbf{\tilde e}_{N}(Z)$ for {\it
  any} $A$ whereas the reverse is true in the special case that $A$
is the refinement of an aggregate assertion $\tilde A$.
Equivalence of the above events implies that the $(p,q,r)$ output for $A$
based on model (\ref{eq:ddsm1}) equals the $(p,q,r)$ output for $\tilde A$
based on model (\ref{eq:ddsm2}), regarded as the \dDSM for aggregate
data $\tilde N$.
\eop

\section{Calculations for trinomial inference}
\label{sec:calc-trin-infer}

\subsection{Dirichlet-DSM}
The posterior distribution for $\theta$ is represented by the random set,
\[
\mathbf{e}_N(Z) = \{\theta: \theta=(\theta_1,\theta_2,\theta_3) \in
\Theta, \theta_1\geq Z_1, \theta_2\geq Z_2, \theta_3\geq Z_3\},
\]
where $(Z_0, Z_1,Z_2, Z_3) \sim \mbox{Dirichlet}(1, N_1, N_2, N_3)$.  Thus, for each of the assertions $\{\theta_k\leq \theta_0\}$ for each $k$ and $\theta_0\in [0, 1]$, the random draw of $(Z_0, Z_1,Z_2, Z_3)$ is evidence for $\{\theta_k\leq \theta_0\}$ if and only if $Z_k +Z_0 \leq \theta_0$.  The random draw of $(Z_0, Z_1,Z_2, Z_3)$ is evidence against $\{\theta_k\leq \theta_0\}$ if and only if $Z_k > \theta_0$.  So we have
\[
p = \Prob{Z_k + Z_0 \leq \theta_0}, ~
q = \Prob{Z_k > \theta_0},
\]
and $r = 1-(p+q)$.  These values are easily computed by noting that $Z_k \sim \mbox{Beta}(N_k, N_{\bullet}+1-N_k)$ and $Z_k + Z_0 \sim \mbox{Beta}(N_k+1, N_{\bullet}-N_k)$ where $N_{\bullet} = N_1 + N_2 + N_3$.

The assertion $\{\theta_2/\theta_3\leq r_0|\theta_1=0.5\}$ can be rewritten as $\{\theta_3\geq 0.5/(1+r_0)|\theta_1=0.5\}$. Given a draw of $(Z_0,Z_1,Z_2,Z_3)$, the random set for $(\theta_2,\theta_3)$ obtained by intersecting the random simplex with $\theta_1=0.5$ is not empty only when $Z_1<0.5<Z_0+Z_1$. It has coordinates $(0.5,Z_3+0.5,Z_3)$ and $(0.5,Z_2, Z_2+0.5)$ in the simplex $\mathcal{S}_2$.  Therefore, the random set for $\{\theta_3|\theta_1=0.5\}$ is $(Z_3,Z_2+0.5)$ given that $Z_1<0.5<Z_0+Z_1$. Thus, DS-$(p,q,r)$ can be easily calculated as
\[
p=Pr\big(Z_3\geq 0.5/(1+r_0)|Z_1<0.5<Z_0+Z_1\big),
\]
\[
q=Pr\big(Z_2+0.5\leq 0.5/(1+r_0)|Z_1<0.5<Z_0+Z_1\big)
\]
and $r=1-p-q$.

In the case when $N=(1,0,0)$, we have DS-$(p,q,r)=(0,0,1)$ by noticing that $Z_2=Z_3=0$ with probability one.

\subsection{Simplex-DSM}
\cite{dempster1968} and \cite{dempster1972} discuss computational issues with the $\sDSM$. He derives the distributions of vertices with minimum and maximum $U_i$s, where $U_i$s are the $i$-th coordinates of the vertices in the random simplex.  His result leads to closed-form solution to the calculation of DS$(p,q,r)$ for simple assertions such as $\{\theta_k\leq\theta_0\}$ for $k=1,2,3$, but not for general assertion such as the conditional assertion. Here, we use a Monte Carlo approach to estimate the triplet $(p,q,r)$ for each assertion based on a sample of random polytopes.  Since our data sets are small, we develop an acceptance-rejection algorithm to compute the random set and the DS$(p,q,r)$ assessments for the assertions of interest.

For simplicity, consider the case that the individual observations is represented by $X_1=1, X_2=1, X_3=2$ and $X_4=3$, which produces the table of counts $N=(2,1,1)$. The random set for $\theta$ given the observed data $N=(2,1,1)$ is the non-empty intersection of $4=N_1+N_2+N_3$ polygons defined by the following set of constraints:
\begin{equation}
     \frac{Z_{i,X_i}}{\theta_{X_i}} \leq \frac{Z_{i,j}}{\theta_{j}}
             \qquad (j\neq X_i; i=1,\ldots,4),
\label{eq:sDSM5.1.1}
\end{equation}
where $Z_{i,j}$ are IID Expo(1) for $i=1,\ldots,4$ and j=1,2,and 3. The acceptance-rejection algorithm generates a draw of the random set as follows:

{\em Step 1. Draw $Z_{i,j}$ from Expo(1) for $i=1,\ldots,4$ and j=1,2,and 3.}

{\em Step 2. Return $Z_{i,j}$ if there exists at least one $\theta\in \Theta$ such as (\ref{eq:sDSM5.1.1}) holds, and go to Step 1 otherwise.}

We make use of the linear programming for Step 2 by solving the optimization problem with the objective functon 
\[
f(\theta)=\theta_1
\]
subject to the constraints (\ref{eq:sDSM5.1.1}) which can be written as the linear constraints:
\begin{equation}
     Z_{i,X_i}\theta_j-Z_{i,j}\theta_{X_i}\leq 0 \qquad (j\neq X_i; i=1,\ldots,4)
\label{eq:sDSM5.1.2}
\end{equation}
in addition to the simplex constraint $\theta\in \Theta$.

The DS$(p,q,r)$ for the conditional assertion $\{\theta_2/\theta_3\leq r_0|\theta_1=0.5\}$ can be computed by find the extreme values of the objective function 
\[
f(\theta)=\theta_2-r_0\theta_3
\]
subject to the constraints in (\ref{eq:sDSM5.1.2}) and $\theta_1=0.5$. Let $[A,B]$ be the random interval for the objective function. Then  the case of $A>0$ provides an instance against for the assertion. The case of $B\leq 0$ provides an instance for the assertions. The remaining case $A\leq 0 <B$ contributed to the probability of ``don't know".

\subsection{IDM}
The IDM consists of a family of Bayesian priors
\[
  \{\mbox{$\prod_{k=1}^K \theta_k ^{\alpha_k-1}: \sum_{k=1}^K \alpha_i \leq s$}\},
\]
which results in the family of posterior distributions
\begin{equation}
  \{\mbox{$\prod_{k=1}^K \theta_k ^{N_k+\alpha_k-1}: \sum_{k=1}^K \alpha_i \leq s$}\}.
\label{eq:IDMposterior}
\end{equation}
The probability $p_{\alpha}$ for the assertion $\{\theta_k\leq \theta_0\}$ is
\[
p_\alpha = \Prob{\theta_k < \theta_0}, \qquad \theta_k \sim \mbox{Beta}(N_k+\alpha_k, N_{\bullet}+s-N_k-\alpha_k)
\]
Take $s=1$.  Thus the lower probability $p$ is obtained by taking $\alpha_k = 1$, that is,
\[
p = \Prob{\theta_k < \theta_0}, \qquad \theta_k \sim \mbox{Beta}(N_k+1, N_{\bullet}-N_k)
\]
which is the same as that based on $\dDSM$.  The upper probability $p+r$ is also the same as that based on $\dDSM$:
\[
p+r =  \Prob{\theta_k < \theta_0}, \qquad \theta_k \sim \mbox{Beta}(N_k, N_{\bullet}+1-N_k)
\]

For the Bayesian inference, we consider Jeffreys' prior which is
equivalent to a single distribution from the IDM model with $\alpha_k
= 1/2$ for all $k$ and $s = K/2$.  As this yields only a single
probability, we have $r = 0$.

\section{An imprecise Dirichlet model for the linkage example}
\label{sec:linkage}
This section describes the IDM linkage analysis.

Begin with the IDM family of priors for a $K=4$, multinomial analysis:
\begin{equation*}
\pi(\theta | \alpha) \propto \theta_{1}^{\alpha_1 - 1} \theta_{2}^{\alpha_2 - 1} \theta_{3}^{\alpha_3 - 1} (1 - \theta_1 - \theta_2 - \theta_3)^{\alpha_4 - 1},
\end{equation*}
where $\sum_{k=1}^4 \alpha_i = 1$.
Consider the following transformation
\begin{eqnarray*}
\phi & = & 4(1 - \theta_1 - \theta_2 - \theta_3) \\
\eta_1 & = & \theta_2 - \frac{1}{4} + (1 - \theta_1 - \theta_2 - \theta_3) \\
\eta_2 & = & \theta_3 - \frac{1}{4} + (1 - \theta_1 - \theta_2 - \theta_3),
\end{eqnarray*}
where $\phi$ is the desired parameter and $\eta_1$ and $\eta_2$ are
nuisance parameters that the linkage model takes to be zero.  The
Jacobian for this transformation is a
constant.  The family of IDM priors in terms of these parameters is
\begin{equation*}
\pi(\phi, \eta_1, \eta_2) \propto \left(\frac{1}{2} - \eta_1 - \eta_2 + \frac{\phi}{4}\right)^{\alpha_1 - 1} \left(\eta_1 + \frac{1}{4} - \frac{\phi}{4}\right)^{\alpha_2 - 1} \left(\eta_2 + \frac{1}{4} - \frac{\phi}{4}\right)^{\alpha_3 - 1} \left(\frac{\phi}{4}\right)^{\alpha_4 - 1}
\end{equation*}
Conditioning on $\eta_1 = \eta_2 = 0$ imposes the desired one-dimensional constraint and gives
\begin{equation*}
\pi(\phi | \eta_1=0, \eta_2=0) \propto \left( \frac{2+\phi}{4} \right)^{\alpha_1 - 1} \left( \frac{1-\phi}{4} \right)^{\alpha_2 + \alpha_3 - 1} \left(\frac{\phi}{4} \right)^{\alpha_4 - 1},
\end{equation*}
with $\sum_{k=1}^4 \alpha_i = 1$.

The resulting family of posteriors can be explored using MCMC over a
grid of a $\alpha$ values that satisfy the
constraint.  In this case, the bounding cases appear to correspond to
$\alpha_4 = 1$ on the lower side and $\alpha_2 = 1$ or $\alpha_3=1$ on
the upper side.

\end{document}